\documentclass[journal]{IEEEtran}

\usepackage{amsmath,amsfonts,bm}
\usepackage{algorithmic}
\usepackage{array}
\usepackage[caption=false,font=normalsize,labelfont=sf,textfont=sf]{subfig}
\usepackage{textcomp}
\usepackage{stfloats}
\usepackage{url}
\usepackage{verbatim}
\usepackage{graphicx}

\usepackage{textcomp}
\usepackage[dvipsnames]{xcolor}
\usepackage{colortbl}
\definecolor{tblhead}{HTML}{DCE6F2}
\definecolor{tblgroup}{HTML}{F2F2F2}
\definecolor{tblstripe}{HTML}{F7F9FC}
\definecolor{tblprimary}{HTML}{EAF2FB}
\definecolor{passbg}{HTML}{E3F4E1}
\definecolor{passfg}{HTML}{1E7B34}
\definecolor{partbg}{HTML}{FFF1D6}
\definecolor{partfg}{HTML}{A15C00}
\definecolor{failbg}{HTML}{FBE1E1}
\definecolor{failfg}{HTML}{B22222}
\newcommand{\tpass}[1]{\cellcolor{passbg}\textcolor{passfg}{\ding{51}~#1}}
\newcommand{\tpart}[1]{\cellcolor{partbg}\textcolor{partfg}{$\sim$~#1}}
\newcommand{\tfail}[1]{\cellcolor{failbg}\textcolor{failfg}{\ding{55}~#1}}
\newcommand{\tgroup}[2]{\rowcolor{tblgroup}\multicolumn{#1}{l}{\textbf{\textit{#2}}}\\}
\usepackage[linesnumbered,ruled,vlined]{algorithm2e}
\usepackage{hyperref}
\usepackage{tabulary}

\usepackage{float}

\usepackage{graphics}
\graphicspath{ {./figs/} }

\usepackage{caption}
\usepackage{subcaption}

\usepackage{wrapfig}

\usepackage[nobiblatex]{xurl}

\usepackage{amsthm}
\newtheoremstyle{mystyle} %
  {3pt}   %
  {3pt}
  {} %
  {}      %
  {} %
  {}     %
  { }     %
  {}      %

\theoremstyle{mystyle}

\newtheorem{metric}{Metric}

\usepackage{booktabs}

\usepackage{tabularx}
\usepackage{ragged2e} %

\usepackage{makecell}
\usepackage{multirow}

\usepackage{paralist}

\usepackage{booktabs}
\usepackage{siunitx}

\usepackage{mathtools,tikz,caption}
\DeclareRobustCommand\sampleline[1]{%
    \tikz\draw[#1] (0,0) (0,\the\dimexpr\fontdimen22\textfont2\relax)
    -- (2em,\the\dimexpr\fontdimen22\textfont2\relax);%
}

\usepackage{pifont}
\usepackage[most]{tcolorbox}
\tcbset{
  colback=green!5!white,
  colframe=green!85!black,
  boxsep=4pt,                 %
  left=0pt, right=0pt,        %
  top=0pt, bottom=0pt,        %
  enhanced,
  before skip=2pt,
  after skip=3pt,
  sharp corners               %
}

\usepackage{soul}
\newcommand{\step}[1]{%
  \tikz[baseline=(char.base)]{
    \node[shape=circle,fill=black,inner sep=0.8pt] (char) {\textcolor{white}{#1}};
  }%
}

\newcounter{rqcounter}
\newcounter{subrqcounter}[rqcounter]
\newcommand{\newrq}[2]{%
    \begin{description}
    \refstepcounter{rqcounter}%
    \setcounter{subrqcounter}{0}%
    \item[\textbf{RQ\arabic{rqcounter}:}] {\em #2}\label{#1}%
    \end{description}
}

\newcommand{\rqref}[1]{\textbf{RQ\ref{#1}}}

\definecolor{color-up-to-date}{HTML}{8CC5E3}
\definecolor{color-tood}{HTML}{2066A8}
\definecolor{color-pfet-tood}{HTML}{F72B8F}

\usepackage{lipsum}                     %
\usepackage{xargs}                      %
\usepackage{enumitem}
\setlist[enumerate]{nosep}

\usepackage[colorinlistoftodos,prependcaption,textsize=scriptsize]{todonotes}
\newcommandx{\unsure}[2][1=]{\todo[linecolor=red,backgroundcolor=red!25,bordercolor=red,#1]{#2}}
\newcommandx{\change}[2][1=]{\todo[linecolor=blue,backgroundcolor=blue!25,bordercolor=blue,#1]{#2}}
\newcommandx{\info}[2][1=]{\todo[linecolor=OliveGreen,backgroundcolor=OliveGreen!25,bordercolor=OliveGreen,#1]{#2}}
\newcommandx{\improvement}[2][1=]{\todo[linecolor=Plum,backgroundcolor=Plum!25,bordercolor=Plum,#1]{#2}}
\newcommandx{\thiswillnotshow}[2][1=]{\todo[disable,#1]{#2}}

\SetCommentSty{mycommfont}

\newcommand{\pkgdep}{{$<$package, dependency$>$} }

\newcommand{\mttu}{MTTU\textsubscript{dep}\xspace}
\newcommand{\mttr}{MTTR\textsubscript{dep}\xspace}
\newcommand{\MTTU}{Mean-Time-To-Update\textsubscript{dep}\xspace}

\newcommand{\mttuw}{MTTU\textsubscript{dep}\xspace}
\newcommand{\mttrw}{MTTR\textsubscript{dep}\xspace}

\newcommand{\ttu}{TTU\textsubscript{dep}\xspace}
\newcommand{\ttr}{TTR\textsubscript{dep}\xspace}

\newcommand\pkgname[1]{\textsf{\small #1}}
\newcommand\version[1]{$\operatorname{#1}$}

\newcommand\goalstatement{\textit{The goal of this study is to aid practitioners
in understanding how quickly packages update their dependencies
through an empirical study using two novel dependency update metrics.
}}

\begin{document}

\title{How Quickly Do Development Teams Update Their Vulnerable Dependencies?}

\author{
Imranur~Rahman$^{*}$, Ranindya~Paramitha$^{\dagger}$, Nusrat~Zahan$^{\ddagger}$, Jason~Osborne$^{*}$, William~Enck$^{*}$, Laurie~Williams$^{*}$
\\ $^{*}$North Carolina State University, \{irahman3, jaosborn, whenck, lawilli3\}@ncsu.edu
\\ $^{\dagger}$University of Bristol, ranindya.paramitha@bristol.ac.uk
\\ $^{\ddagger}$Columbia University, nz2479@columbia.edu
}

\maketitle

\begin{abstract}
Industry practitioners are increasingly concerned with software that contains vulnerable versions of third-party dependencies that are included both directly and transitively.
To address this problem, projects are encouraged to both (a)~quickly update to non-vulnerable versions of dependencies and (b)~be mindful of the update practices of the dependencies they choose to use.
To this end, researchers have proposed metrics to measure the responsiveness of the development teams of the packages in keeping their dependencies updated: Mean-Time-To-Update (MTTU) and Mean-Time-To-Remediate (MTTR).
While MTTU covers all dependencies, MTTR quantifies the time needed for a package to update its vulnerable dependencies.
However, existing metrics fail to capture important nuances, such as considering floating versions and prioritizing recent updates, leading to inaccurate reflections of a development team's update practices.
\textit{The goal of this study is to aid practitioners in understanding how quickly packages update their dependencies through an empirical study.}
We propose two novel metrics, Mean-Time-To-Update for dependencies (\mttu) and Mean-Time-To-Remediate for vulnerable dependencies (\mttr), that overcome the limitations of existing metrics.
We conduct an empirical study using $163{,}207$ packages in npm ($117{,}129$), PyPI ($42{,}777$), and Cargo ($3{,}301$) and 
characterize how the ecosystems differ in \mttu and \mttr, as well as what package characteristics influence \mttu and \mttr.
We found that most maintained packages have a relatively fast dependency update practice.
We further study whether \mttu can serve as a proxy for \mttr when sufficient vulnerability data are not available.
After controlling for nine characteristics of the package, its repository, and its developers, \mttu still explains 28.7\% of the variance in \mttr that these characteristics leave unexplained.
Thus, \mttu can serve as a context-adjusted proxy that shows which packages are likely to remediate slowly, although it does not replace \mttr.
This latter finding is particularly important given that only $1{,}363$ npm (0.04\%), 694 PyPI (0.11\%), and 383 Cargo (0.20\%) packages have reported vulnerabilities, and the existence of \mttu will allow practitioners to make more informed decisions about the dependencies they choose.
\end{abstract}


\section{Introduction}
Vulnerable dependencies are widely present in both open-source software (OSS) and proprietary codebases.
According to the Black Duck 2025 ``Open Source Security and Risk Analysis Report''~\cite{blackduck}, 86\% of codebases contain at least one vulnerable open source dependency, and 81\% of codebases contain high or critical risk externally reported vulnerabilities resulting from dependencies.
The majority of codebases had vulnerable dependencies for more than two years despite the availability of a fixed version~\cite{blackduck-2021}.
This delay occurs because of fear of breaking changes and the cost associated with updating vulnerable dependencies to fixed versions~\cite{derrKeepMeUpdated2017}.

This issue of vulnerable dependencies highlights the industry practitioners' need for metrics to measure the responsiveness of development teams to updating their open-source dependencies.
For example, the OpenSSF Scorecard~\cite{ossf-scorecard}
evaluates a package based on 18 security practices.
Among these practices are the ``maintained'' check, which determines if the project is maintained by checking activity in the last 90 days, and the ``vulnerabilities'' check, which detects if there are unfixed externally reported vulnerabilities in the project or its dependencies.

While Scorecard's ``maintained'' metric focuses on recent activity, other metrics, such as Mean-Time-To-Update (MTTU) and Mean-Time-To-Remediate (MTTR), provide a historical perspective on how long it takes for the development team of a package to update their dependencies.
Industry reports introduced these two metrics for dependencies~\cite{sonatype-2019}.
MTTU captures the time to update all dependency versions, while MTTR focuses specifically on the time to update vulnerable dependency versions to the fixed version.
MTTR has been extensively discussed in the context of a project fixing its own vulnerabilities, but not vulnerabilities in dependencies~\cite{gokhaleTimeStructureBased1999,calvoApplyingGoalQuestion2023,johnsonSurveySoftwareTools1988,morrisonMappingFieldSoftware2018}.
Dependency update metrics, such as technical lag~\cite{gonzalez-barahona_technical_2017}, consider management of dependency versions; however, they have not been applied to vulnerabilities.
Furthermore, active maintenance (e.g., regular dependency update) is a desired criterion for developers in selecting a dependency as highlighted by Vargas et al.~\cite{larios_vargas_selecting_2020}.
In this context, a lower value in a dependency update metric suggests faster and more consistent updates, signaling ongoing maintenance and reliability~\cite{sonatype-2024}.

Practitioners measuring the responsiveness of a team in updating its vulnerable and outdated dependencies using dependency update metrics face two key challenges.

\noindent
\textbf{1. Metric Limitations.} 
The calculations used by existing dependency update metrics often do not handle floating version constraints or are not designed to localize vulnerabilities~\cite{cox_measuring_2015,gonzalez-barahona_technical_2017}.
Existing metrics also often calculate an update time \emph{for each} of a dependency and \emph{for each} version of a package, making it difficult to understand that package's update practices as a whole.
Additionally, existing metrics do not weight the recent dependency update practice which makes the existing metrics less actionable for developers.
Finally, the existing metrics for vulnerable dependencies, such as vulnerability lag~\cite{zeroualiMultidimensionalAnalysisTechnical2021}, are computed for each version or each vulnerability, and do not give one value for the remediation practice of a package.
Some industry reports use MTTR for this purpose, but their specific calculations are not available.

\noindent\textbf{2. Insufficient Externally Reported Vulnerability Data.}
A small fraction of packages in software ecosystems have reported vulnerabilities.
For example, 1,363 npm, 694 PyPI, and 383 Cargo (total 2.4K) packages have externally-reported vulnerabilities as of 2026-05-07.
Practitioners can only compute MTTR for the dependents of those 2.4K packages.
0.49\% npm, 0.83\% PyPI, and 0.05\% Cargo packages have directly depended on at least one of these 2.4k vulnerable packages in their lifetime.
In addition, the lack of externally reported vulnerability data is also discussed by related research~\cite{alhazmi2007measuring,zahan_software_2023}.

\goalstatement
With this goal, we conduct our study with four research questions. We split two of them (RQ2 and RQ3) into two sub-questions each.

\noindent \newrq{rq:metrics}{How do we measure the MTTU and MTTR of a package including its dependencies such that the measure is responsive to more recent update practices and therefore actionable to the development team?}
We first propose novel algorithms for computing Mean-Time-To-Update (MTTU) and Mean-Time-To-Remediate (MTTR), which we denote \mttu and \mttr, respectively, to overcome challenges with existing dependency update metric calculations.

After defining the two novel metrics, we performed an empirical analysis on npm, PyPI, and Cargo ecosystems.
We collect package version release information and dependency relations of packages of $163{,}207$ packages from the three ecosystems and compute \mttu and \mttr.
With the computed metrics, we answer the second research question.

\noindent \newrq{rq:mttu-mttr}{How do packages in npm, PyPI, and Cargo differ in MTTU and MTTR?}
We analyze the distributions of \mttu and \mttr using violin plots, then test the ecosystem differences directly with Kruskal--Wallis, Mann--Whitney, and Cliff's delta (RQ2a), and test whether remediation speed depends on vulnerability severity (RQ2b).

In all three ecosystems, \mttr cannot be computed for 99.51\% npm, 99.17\% PyPI, and 99.95\% Cargo packages since these packages have not depended upon any vulnerable direct dependencies in their lifetime. While dependency update metrics are useful, this lack of externally reported vulnerability data limits their application. 
To provide dependency update measurements for those 99\%+ of packages without externally-reported vulnerabilities, we analyzed whether \mttu would provide a practical estimation/ proxy. This analysis is inspired by Dhrymes and Guerard~\cite{dhrymes1978introductory}, who suggested that when a variable is unobservable, a proxy variable, which is a variable that can be used as a substitute for the missing one, can be used.
Our aim is not to predict the exact \mttr value of a package from its \mttu value.
Our aim is to estimate how a package is likely to remediate vulnerable dependencies from behavior that we can observe for every package, read in the context of the package, its repository, and its developers.
This leads us to our third research question.

\noindent \newrq{rq:substitute}{Can MTTU be used as a proxy for MTTR?}
We perform a proxy analysis, using analyses informed by prior work, on \mttu and \mttr to explore whether \mttu can serve as a proxy for \mttr (RQ3a).
Both metrics can depend on the same package context, such as size, popularity, and activity.
We therefore control for nine package characteristics in the proxy analysis, so that \mttu counts as a proxy only if it tells us something about \mttr that the context does not already tell us.
We then recompute this proxy analysis with varying weighting functions (no weighting and inverse, in addition to our primary exponential weighting) to understand the sensitivity of our weighting mechanism on our conclusions (RQ3b).

Finally, we would like to understand which package characteristics (e.g., contributors count, version count) have more influence on \mttu and \mttr, which leads to our last research question.

\noindent \newrq{rq:characteristics}{How do package characteristics influence MTTU and MTTR?}
We use correlation tests to understand the association between nine package characteristics and \mttu/ \mttr.
We further conduct regression analysis to quantify which package characteristics matter more in influencing \mttu and \mttr values.

\noindent
\textbf{Contributions.}
In summary, this paper contributes 
(1)~a detailed algorithm and process for quantifying the dependency update practice of a package using our novel metrics; 
(2)~a large-scale analysis of the dependency update metrics in npm, PyPI, and Cargo packages, including statistical tests of how the ecosystems differ and whether remediation speed depends on vulnerability severity;
(3)~a covariate-adjusted proxy analysis showing that MTTU remains informative about MTTR after controlling for nine package characteristics, for use when externally reported vulnerability data is not available, together with a weighting sensitivity analysis showing which of these proxy-analysis conclusions hold regardless of the metrics' weighting function, and which depend on it; and
(4)~correlation and regression analysis illustrating which package characteristics impact the likelihood of higher MTTU and MTTR.

We provide our replication package in Zenodo~\cite{zenodo-artifact}, currently restricted for reviewers only.
Upon acceptance of the paper, we will make it public.

\section{Background And Related Work}
\label{sec:rel-work}
In this section, we provide a brief overview of the existing update metric and discuss related work.

\textbf{Technical Lag.} %
The concept of \textit{technical lag} was first introduced by Gonzalez-Barahona et al.~\cite{gonzalez-barahona_technical_2017} for OSS packages.
Essentially, ``technical lag'' quantifies how quickly software systems fall behind as new versions and updates are released.
Zerouali et al.~\cite{zerouali_empirical_2018} applied ``technical lag'' to the context of dependencies and conducted an empirical analysis of package dependency updates in the npm ecosystem.
They found that outdated dependencies induce a median technical lag of 3.5 months in npm.
Building on this, Decan et al.~\cite{decan_evolution_2018} conducted a longitudinal empirical study of `technical lag' in the npm dependency network and explored how technical lag increases over time.
They observed that technical lag for most npm packages increases during their lifespan, and technical lag occurred mainly due to the minor and patch releases of a dependency.

Further research by Zerouali et al.~\cite{zeroualiFormalFrameworkMeasuring2019} propose a formal framework for measuring technical lag in software ecosystems.
They analyzed 4M releases of 500K npm packages, considering the evolution of technical lag over time.
They found that technical lag induced by direct dependencies in npm packages increases over time due to missed updates, including major releases.
Stringer et al.~\cite{stringerTechnicalLagDependencies2020} study the technical lag of dependencies in a large-scale cross-ecosystem fashion containing packages from 14 package managers.
They found that pinned dependencies are the main reason behind technical lag.
Zerouali et al.~\cite{zeroualiMultidimensionalAnalysisTechnical2021} expand the idea of `technical lag' into multiple dimensions (package lag, time lag, version lag, vulnerability lag, and bug lag) and study the technical lag in 140K Docker images.
They found that the median time lag of community Docker images is over a year.
Although previous studies have explored technical lag in terms of general dependency updates, understanding vulnerable dependency update practice was not their goal.
In contrast, we study the dependency update metric and the vulnerable dependency update metric and investigate their relationship and other characteristics.

\textbf{Outdated and vulnerable dependencies.}
Kula et al.~\cite{kulaTrustingLibraryStudy2015} analyze the latency in adopting the latest version of a dependency in the Maven ecosystem.
Their study found that developers are more likely to adopt the latest version for newly added dependencies than existing ones.
In a follow-up study, Kula et al.~\cite{kula_developers_2018} examined library migration across GitHub projects and found that 81.5\% of the projects keep using outdated dependencies.
Cox et al.~\cite{cox_measuring_2015} introduce the concept of `dependency freshness' to study dependency updatedness in the Maven ecosystem.
They found that only $16.7\%$ of the dependencies display no update lag.
Derr et al.~\cite{derrKeepMeUpdated2017} identify the root causes of outdated dependencies in the Android ecosystem and find that developers do not update their third-party library dependencies due to fear of breaking changes, lack of knowledge, and lack of motivation.
Wang et al.~\cite{wangEmpiricalStudyUsages2020} conduct an empirical study on dependency update analysis on OSS packages and find that 50\% of the packages use outdated dependencies.
Huang et al.~\cite{huangCharacterizingUsagesUpdates2022} extend Wang et al.~\cite{wangEmpiricalStudyUsages2020}'s study and find that one-third of the projects have a lag of one major version from the latest library version.

Pashchenko et al.~\cite{pashchenkoVulnerableOpenSource2018} studied the most used Java dependencies in SAP software and found that only updating the dependencies' version can remove 81\% vulnerable dependencies.
The same study by Kula et al.~\cite{kula_developers_2018} also examined how developers respond to security advisories, and found that developers often do not update their vulnerable dependencies.
Kumar et al.~\cite{kumar_comprehensive_2024} conducted a study to understand how widespread vulnerabilities are and how quickly they are being fixed.
They found that for most programming languages, a critical vulnerability persists on average for over a year before being fixed.

Several studies measure how long the fix for a vulnerable dependency takes to reach the dependent packages.
Decan et al.~\cite{decan_impact_2018} studied how vulnerabilities in npm are discovered and fixed, and how the dependency constraints of dependent packages affect the adoption of the fixes.
Zerouali et al.~\cite{zerouali_impact_2022} extended this analysis to npm and RubyGems, and found that many dependents are exposed to vulnerabilities whose fixes exist in the same major release range of the dependency.
Chinthanet et al.~\cite{chinthanet_lags_2021} measured the lags in the release, the adoption, and the propagation of npm vulnerability fixes.
Liu et al.~\cite{liuDemystifyingVulnerabilityPropagation2022} resolved complete npm dependency trees and studied how vulnerabilities propagate through them over time.
Alfadel et al.~\cite{alfadel_empirical_2023} found that dependent Python projects take about seven months to update a vulnerable package to a non-vulnerable version.
These studies measure lags for each vulnerability or each dependency.
In contrast, our metrics give one value per package, so that a practitioner can compare the remediation practice of two packages.

Other studies examine bots that automate dependency updates.
Alfadel et al.~\cite{alfadel_use_2021} studied Dependabot security pull requests and found that developers merge most of them, often within a day.
He et al.~\cite{he_automating_2023} found that Dependabot reduces technical lag, but that developers configure it to reduce its notifications, and 11.3\% of the projects stop using it.
Such bots can change both \mttu and \mttr of a package, so we discuss them as an unmeasured factor in Section~\ref{sec:threats}.

Studies in \textit{outdated dependencies} and \textit{vulnerable dependencies} are focused on either all updates or only security updates, but not both.
In contrast, we focus on both outdated and vulnerable dependencies and explore their relationship using our proposed metrics.

\textbf{MTTU and MTTR.}
The metrics MTTU and MTTR have been used in the software reliability and maintenance domain for a long time~\cite{gokhaleTimeStructureBased1999,calvoApplyingGoalQuestion2023,johnsonSurveySoftwareTools1988,morrisonMappingFieldSoftware2018}.
Researchers have also studied different security metrics (time to close bug/vulnerability, window of exposure, vulnerability count)~\cite{davidsonGoodBadUgly2009,beresUsingSecurityMetrics2009} of various categories (time metric, vulnerability metric) in the software security domain, but these metrics are focused on vulnerabilities in packages\footnote{Faults and bugs are considered in MTTR instead of vulnerability in Reliability domain research.}.
In our study, we focus on measuring dependency update metrics for packages having vulnerable dependencies, not the vulnerable package itself.

MTTR has also been used in different contexts in the industry, e.g., measuring the package's security~\cite{mttr-dark-reading,mttr-plextrac} and measuring the package's security in terms of dependency~\cite{sonatype-2019}.
The procedures for measuring MTTR and MTTU are often proprietary and not disclosed for academic research.
For example, Sonatype's 2024 report~\cite{sonatype-2024} measured MTTU and MTTR for Maven, but we could not reproduce it since the methodology is not available.

\section{Challenges Applying Existing Update Metrics}
\label{sec:challenges}
In this section, we first provide an example of how the most prominent update metric with a published algorithm available in the literature, technical lag, works.
Then, we describe the design gaps present in technical lag when measuring the updatedness of dependencies and why new metrics are needed.

\begin{table}[ht]
    \footnotesize
    \centering
    \caption{tLag measurement of \pkgname{codemod-cli} package with one of its dependency, \pkgname{simple-git}.
    }
    \resizebox{\linewidth}{!}{
    \begin{tabular}{c|c|r|c|r}
    \toprule
    \makecell[t]{codemod-cli\\\emph{version: date}} & \makecell[t]{simple-git\\constraint} & \makecell[t]{lastAllowed(simple-git)\\\emph{version: date}} & \makecell[t]{latest(simple-git)\\\emph{version: date}} & \makecell[t]{tLag}
    \\
    \midrule
    $0.8.6 : \operatorname{2022-03-03}$ & $^\wedge1.130.0$ & $1.132.0 : \operatorname{2020-03-12}$ & $3.2.6 : \operatorname{2022-02-17}$ & $707$
     \\ \hline
    $0.8.7 : \operatorname{2022-03-05}$ & $^\wedge2.48.0$ & $2.48.0 : \operatorname{2021-12-01}$ & $3.2.6 : \operatorname{2022-02-17}$ & $78$
     \\ \hline
    $0.9.0 : \operatorname{2022-03-08}$ & $^\wedge2.48.0$ & $2.48.0 : \operatorname{2021-12-01}$ & $3.2.6 : \operatorname{2022-02-17}$ & $78$ 
     \\ \hline
    $0.9.1 : \operatorname{2022-03-17}$ & $^\wedge2.48.0$ & $2.48.0 : \operatorname{2021-12-01}$ & $3.3.0 : \operatorname{2022-03-11}$ & $100$
     \\ \hline
    $0.9.2 : \operatorname{2022-03-22}$ & $^\wedge3.4.0$ & $3.4.0 : \operatorname{2022-03-18}$ & $3.4.0 : \operatorname{2022-03-18}$ & $0$
     \\ \bottomrule
    
    \end {tabular}
    }
    \label{table:prev-works}
\end{table}

\subsection{Technical Lag Using \emph{tLag}}

Time lag ($tLag$)~\cite{zerouali_empirical_2018,decan_evolution_2018} is a way to measure technical lag to assess the outdatedness of a dependency in terms of time.
Conceptually, $tLag$ measures the time difference between the release time of the latest version of a dependency and the release time of the version of the dependency used by a package, at the time the package was released.

Let $pkg$ be a package, and $pkg_v$ indicate a specific version $v$ of $pkg$.
Let $dep$ be a direct dependency of $pkg$.
When specifying $dep$ as a dependency, $pkg_v$ specifies a dependency constraint $pkg_{v}.dep.\bar{c}$.
Package managers use dependency constraints to resolve the highest version that satisfies the constraint.
For example, a dependency constraint $pkg_{v}.dep.\bar{c}$ of ``$^\wedge1.2.3$'' is satisfied for the range [$>=1.2.3, < 2.0.0$].
We use the function $lastAllowed(dep, \bar{c}, t)$ to denote the version resolution of $dep$ for constraint $\bar{c}$ at time $t$.
We further use the function $latest(dep, t)$ to denote the latest version of $dep$ at time $t$ and $time(\cdot)$ to denote the time a given version is released.
Hence, the $tLag$ of $pkg_v$ for $dep$ is formally defined in Equation~\ref{eq:tlag}.

\setlength{\belowdisplayskip}{2pt} \setlength{\belowdisplayshortskip}{0pt}
\setlength{\abovedisplayskip}{-12pt} \setlength{\abovedisplayshortskip}{-5pt}
\begin{multline}
    tLag(pkg_v,dep) = time(latest(dep, time(pkg_{v}))) \\ - time(lastAllowed(dep, pkg_{v}.dep.\bar{c}, time(pkg_{v})))
\label{eq:tlag}
\end{multline}

\begin{figure}
    \centering
    \includegraphics[width=.8\linewidth]{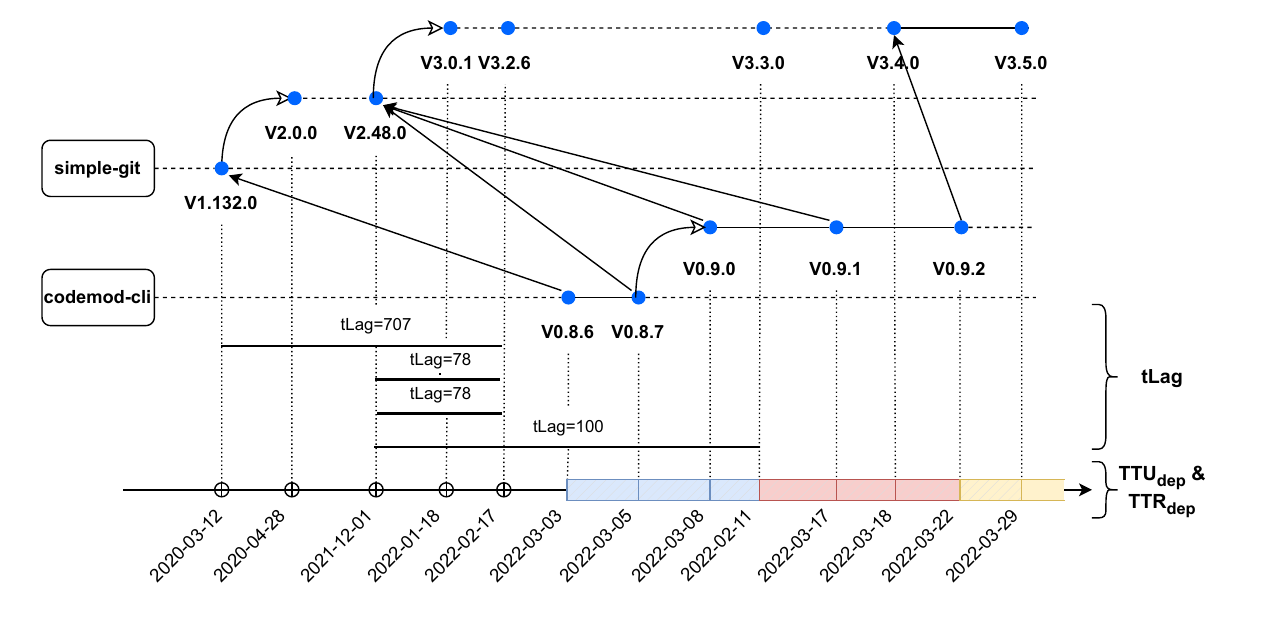}
    \caption{Illustration of tLag, and \mttu/\mttr calculation using \pkgname{codemod-cli}'s dependency relationship with \pkgname{simple-git}. \textcolor{cyan}{Blue} time quantum indicates outdated dependency, \textcolor{red}{red} time quantum indicates outdated and vulnerable dependency, and \textcolor{yellow}{yellow} quantum indicates updated dependency.}
    \label{fig:mttu-mttr-tlag}
\end{figure}

Table~\ref{table:prev-works} shows an example computation of technical lag for \pkgname{codemod-cli}'s dependency on \pkgname{simple-git}.
The first two columns of Table~\ref{table:prev-works} indicate the version with the date of different releases of package \pkgname{codemod-cli} and the version constraint specified for dependency \pkgname{simple-git}.
The ``lastAllowed(simple-git)'' column represents the resolved version of dependency \pkgname{simple-git} with specified constraint \textit{at the time of the release of} \pkgname{codemod-cli}.
For computing the technical lag, we need to know the latest available version of the dependency \pkgname{simple-git} at the time of the release of each version of \pkgname{codemod-cli} according to Eq.~\ref{eq:tlag}, which is represented in column ``latest(simple-git)''.
The \textit{tLag} column is the computed technical lag.
We subtract the release date of lastAllowed(simple-git) from the release date of latest(simple-git) to obtain the technical lag for each release of \pkgname{codemod-cli}.
The computed \emph{tLag} of five versions of \pkgname{codemod-cli} were 707, 78, 78, 100, and 0, consecutively, in Table~\ref{table:prev-works} and Fig.~\ref{fig:mttu-mttr-tlag}.

\subsection{Design Gaps}
\label{sec:gaps}
In this section, we analyze the gaps available in \emph{tLag}, which are used as motivators of the design choice of our proposed metrics.

\noindent \step{1} \textit{Handling Floating Version Constraints.}
Floating version constraints are available in all major OSS ecosystems and are 
 considered a good practice~\cite{he_pinning_2025,jafari_dependency_2023}, as a package gets automatic updates whenever a newer version of the dependency is released.
However, \textit{tLag} does not sufficiently account for automatic updates that are allowed by floating version constraints.
For example, \textit{tLag} only checks the latest available version of the dependency against the package's used dependency version \textit{when} the package releases a new version.
Even if the package releases a new version, and after that, the dependency releases a newer version that can be auto-updated by the package constraint, \emph{tLag} cannot model this case in its design.
Accounting for floating labels at the time of a vulnerability fix is particularly important when considering exposure to vulnerabilities, as an auto-update may automatically remediate a vulnerability.

\noindent \step{2} \textit{Package-Level Metric.}
A dependency update metric should combine all data points into \emph{one} single value per package
to make it easier for developers to understand and compare metrics.
However, \textit{tLag} computes the technical lag for each version of the package for a dependency as shown in Table~\ref{table:prev-works}.
For developers, interpreting and comparing all of these \textit{tLag} values for a package with multiple releases and multiple dependencies is difficult.
Having an aggregation 
would improve the utility of dependency update metrics.

\noindent \step{3} \textit{Localizing Vulnerabilities.}
A security-oriented dependency update metric should incorporate the update characteristics of vulnerable dependencies.
\textit{tLag} includes bug fixes, feature updates, and security fixes as a whole.
Technically, \textit{tLag}'s measurement can be modified to only consider the dependency's security fixes adopted by the package in its formal framework.
Zerouali et al.~\cite{zeroualiMultidimensionalAnalysisTechnical2021} took a step in this direction with \emph{vulnerability lag}, which counts the vulnerabilities that a newer version would fix.
However, vulnerability lag counts vulnerabilities for each version, and it does not measure how long a package takes to adopt a fixed version.

\noindent \step{4} \textit{Weighting Lifetime And Recent Practices.}
A dependency update metric should reflect more heavily the most recent dependency updates made by the package developers.
However, \textit{tLag} does not weight recency into its measurement.
Recent update practice provides the most relevant information for ongoing maintenance of the package.
For example, the development team's update practice from 10 years ago should not be weighted as heavily as their current update practice.
In addition, recency makes a dependency update metric more actionable since a metric with recency can be used to compare two packages with different lifespans (e.g., two packages with the same functionality, one with a shorter lifespan and the other with a longer lifespan).
Actionability is a desired property for good software metrics~\cite{meneely_validating_metrics_2013}.

We illustrate the comparison of our metrics with existing metrics (e.g., time lag and version lag from technical lag~\cite{zerouali_empirical_2018,decan_evolution_2018}, and dependency freshness using version sequence number or VSN~\cite{cox_measuring_2015}) that account for semantic versioning in Table~\ref{tab:dependency_metrics}.

\begin{table*}[htbp]
    \centering
    \caption{Comparison of Dependency Metrics: Traditional vs. Proposed}
    \label{tab:dependency_metrics}
    \small
    \renewcommand{\arraystretch}{1.4} %
    \begin{tabularx}{\textwidth}{@{} 
        >{\raggedright\bfseries}p{2cm} %
        >{\RaggedRight}X %
        >{\RaggedRight}X %
        >{\RaggedRight}X %
        >{\RaggedRight}X %
        >{\RaggedRight\arraybackslash}X %
    @{}}
        \toprule
        \textbf{Feature} & 
        \textbf{Time Lag ($tLag$)} & 
        \textbf{Version Lag ($vLag$)} & 
        \textbf{Dependency Freshness (VSN)} & 
        \textbf{$MTTU_{dep}$} & 
        \textbf{$MTTR_{dep}$} \\
        \midrule
        
        \textbf{Primary Focus} & 
        Temporal distance between the used and latest version. & 
        Semantic distance ($\Delta$Major, $\Delta$Minor, $\Delta$Patch). & 
        Distance in terms of the number of releases missed. & 
        The weighted time a package uses an \textit{outdated} dependency. & 
        The weighted time a package uses a \textit{vulnerable} dependency. \\
        
        \textbf{Measurement Unit} & 
        Time (e.g., days). & 
        Version Tuple (Major, Minor, Patch). & 
        Integer (Count of releases). & 
        Time (Days, exponentially weighted). & 
        Time (Days, exponentially weighted). \\
        
        \textbf{Recency Weighting} & 
        \textbf{None.} Historical delays are weighted equally to current ones. & 
        \textbf{None.} Counts version distance regardless of release date. & 
        \textbf{None.} Counts missed releases regardless of age. & 
        \textbf{Yes.} Applies weighting function to prioritize recent practices. & 
        \textbf{Yes.} Prioritizes recent remediation practices over historical ones. \\
        
        \textbf{Floating Versions} & 
        \textbf{Poor.} Fails to account for auto-updates allowed by floating constraints. & 
        \textbf{N/A.} Measures static distance between resolved versions. & 
        \textbf{N/A.} Measures static distance between resolved versions. & 
        \textbf{Good.} Splits history into intervals to capture auto-updates. & 
        \textbf{Good.} Captures if a floating constraint automatically remediated a vulnerability. \\
        
        \textbf{Aggregation} & 
        Difficult to aggregate into a single package score. & 
        Difficult due to varying version schemas. & 
        Aggregated via risk profile benchmarks. & 
        \textbf{Package-Level.} Combines all dependency intervals into a single value. & 
        \textbf{Package-Level.} Combines all vulnerability intervals into a single value. \\
        
        \textbf{Security Focus} & 
        \textbf{Low.} General outdatedness. & 
        \textbf{Low.} Semantic distance. & 
        \textbf{Low.} General freshness. & 
        \textbf{Medium.} Measures responsiveness to dependency update; partial proxy for security. & 
        \textbf{High.} Specifically measures responsiveness to update known vulnerable dependencies. \\

        \textbf{Configuration} &
        \textbf{None.} Rigid metric. &
        \textbf{None.} Rigid metric. &
        \textbf{None.} Rigid metric. &
        \textbf{Yes.} Developers can configure the metric by the weighting function and the half life according to their needs. &
        \textbf{Yes.} Developers can configure the metric by weighting function and the half life according to their needs. \\
        
        \bottomrule
    \end{tabularx}
\end{table*}

\section{\rqref{rq:metrics} Novel Dependency Update Metrics}
\label{sec:update-metrics}
In this section, we first describe our design choices that fill the gaps presented in Section~\ref{sec:gaps} and lead to our novel metrics design (\mttu and \mttr) for \rqref{rq:metrics}.
Then, we walk through an illustrative example of computing our metrics.
We then provide a detailed methodology for defining our novel dependency update metrics in Section~\ref{sec:definitions}.

\subsection{Intuition}
We design our metrics to only consider direct dependencies since the package only has control over which version to use for the direct dependencies.
We begin by examining each of the \pkgdep relationships from a temporal perspective.
We then split each \pkgdep relation into multiple intervals based on when the package or the dependency has a newer release (major, minor, or patch release).
Our decision to split into intervals is to capture the benefits of floating version constraints without incentivizing this practice (Gap \step{1}).\footnote{To the best of our knowledge, no academic work quantifies optimal dependency specification (e.g., pinning vs floating version constraints) for balancing security benefits and the cost of maintenance.
Therefore, our design does not discriminate between ways to specify dependency constraints~\cite{he_pinning_2025}.}
The choice of incorporating the dependency version releases into our metrics allows us to capture the case where the fixed dependency version is automatically adopted by floating version constraint.
Then, at each interval, the dependency constraint set by the package for the dependency is resolved with only the versions available at the beginning of the interval.
We do this to make sure that the dependency resolution accounts for the historical version releases of the dependency.

After splitting into intervals, we mark as ``updated''=\textcolor{blue}{false} if the resolved dependency version for that interval does not match the highest available version of the dependency at that time and `true' otherwise.
Similarly, we mark as ``remediated''=\textcolor{red}{false} if the resolved dependency version is vulnerable with a fixed version being available at the beginning of that interval and `true' otherwise (Gap \step{3}).
We then aggregate the ``updated'' and ``remediated'' information of each \pkgdep to compute the \mttuw and \mttrw of each package (Gap \step{2}).
This aggregation involves factoring in how old each interval is and weighting based on that.
With our weighting mechanism, recent intervals get near full weight, and older intervals' weight drops exponentially to zero (Gap \step{4}).

We compute the metrics only from direct dependencies, because developers control only the versions of their direct dependencies.
The version of a transitive dependency is set by its parent in the dependency graph, so developers can change it only through their choice of direct dependencies.
An up-to-date set of direct dependencies gives \mttu and \mttr values of zero, but it does not remediate every transitive vulnerability.
Our metrics therefore measure responsiveness for the factors that developers control, and they do not penalize developers for vulnerable transitive dependencies.
We leave transitive dependencies to future work.

\subsection{An Example With Our Metrics}
\label{sec:running-example}

\begin{table*}[t]
    \centering
    \caption{Running example of \pkgname{codemod-cli} package with one of its dependency \pkgname{simple-git}.
    }
    \resizebox{\linewidth}{!}{
    \begin{tabular}{|c|c|c|c|c|c|c|c|c|c|c|c|c|}
    \hline
    \makecell[t]{row} & \makecell[t]{pkg} & \makecell[t]{pkg\\version} & \makecell[t]{dep} & \makecell[t]{dep\\constraint} & \makecell[t]{dep\\version} & \makecell[t]{dep\\highest rel.} & \textbf{\makecell[t]{Interval start}} & \textbf{\makecell[t]{Interval end}} & \makecell[t]{Age Of Interval}
    & \textit{\textbf{\makecell[t]{updated}}} & \textit{\textbf{\makecell[t]{remediated}}} 
    \\
    \hline\hline
    1 & {codemod-cli} & $0.4.0$ & simple-git & $^\wedge1.130.0$ & $1.132.0$ & $2.13.1$ & $\operatorname{2020-07-16}$ & $\operatorname{2020-07-17}$ & 2120 & \textcolor{blue}{false} & true \\ \hline %
     & {$\ldots$} & $\ldots$ & $\ldots$ & $\ldots$ & $\ldots$ & $\ldots$ & $\ldots$ & $\ldots$ & $\ldots$ & $\ldots$ & $\ldots$  \\ \hline
    82 & {codemod-cli} & $0.8.6$ & simple-git & $^\wedge1.130.0$ & $1.132.0$ & $3.2.6$ & $\operatorname{2022-03-03}$ & $\operatorname{2022-03-05}$ & 1524 & \textcolor{blue}{false} & true \\ \hline %
    83 & {codemod-cli} & $0.8.7$ & simple-git & $^\wedge2.48.0$ & $2.48.0$ & $3.2.6$ & $\operatorname{2022-03-05}$ & $\operatorname{2022-03-08}$ & 1521 & \textcolor{blue}{false} & true \\ \hline %
    84 & {codemod-cli} & $0.9.0$ & simple-git & $^\wedge2.48.0$ & $2.48.0$ & $3.2.6$ & $\operatorname{2022-03-08}$ & $\operatorname{2022-03-11}$ & 1518 & \textcolor{blue}{false} & true \\ \hline %
    85 & {codemod-cli} & $0.9.0$ & simple-git & $^\wedge2.48.0$ & $2.48.0$ & $3.3.0$ & $\operatorname{2022-03-11}$ & $\operatorname{2022-03-17}$ & 1512 & \textcolor{blue}{false} & \textcolor{red}{false} \\ \hline %
    86 & {codemod-cli} & $0.9.1$ & simple-git & $^\wedge2.48.0$ & $2.48.0$ & $3.3.0$ & $\operatorname{2022-03-17}$ & $\operatorname{2022-03-18}$ & 1511 & \textcolor{blue}{false} & \textcolor{red}{false} \\ \hline %
    87 & {codemod-cli} & $0.9.1$ & simple-git & $^\wedge2.48.0$ & $2.48.0$ & $3.4.0$ & $\operatorname{2022-03-18}$ & $\operatorname{2022-03-22}$ & 1507 & \textcolor{blue}{false} & \textcolor{red}{false} \\ \hline %
    88 & {codemod-cli} & $0.9.2$ & simple-git & $^\wedge3.4.0$ & $3.4.0$ & $3.4.0$ & $\operatorname{2022-03-22}$ & $\operatorname{2022-03-29}$ & 1500 & true & true \\ \hline %
    
    {$\ldots$} & $\ldots$ & $\ldots$ & $\ldots$ & $\ldots$ & $\ldots$ & $\ldots$ & $\ldots$ & $\ldots$ & $\ldots$ & $\ldots$ & $\ldots$ \\
    \end {tabular}
    }
    \label{table:running-example}
\end{table*}

In this section, we explain how to calculate the dependency update metrics, Time-To-Update (\ttu) and Time-To-Remediate (\ttr), for a \pkgdep relationship.
As a running example, we consider the dependency relation between \pkgname{codemod-cli} and \pkgname{simple-git}, as shown in Table~\ref{table:running-example}.

We split the dependency relations into multiple intervals based on the release of a new version of the package ``\pkgname{codemod-cli}" or dependency ``\pkgname{simple-git}".
In each interval, for resolving the dependency constraint by the package \pkgname{codemod-cli}, we only consider the dependency versions available at the beginning of the interval (``interval start'').
From $V0.4.0$ to $V0.8.6$ (row 1-82), \pkgname{codemod-cli} has the dependency constraint \pkgname{simple-git} $^\wedge1.130.0$ which resolves into $V1.132.0$.
{simple-git} has the highest release $V2.13.1$ at $\operatorname{2020-07-16}$ (row 1) and $V3.2.6$ at $\operatorname{2022-03-03}$.
As a result, \pkgname{codemod-cli} does not have the highest available release of \pkgname{simple-git} in these intervals (rows 1-82).
In $V0.8.7$, \pkgname{codemod-cli} updated the constraint for \pkgname{simple-git} to $^\wedge2.48.0$.
Even with this update, \pkgname{codemod-cli} has not changed the constraint of \pkgname{simple-git} to the available highest major version 3, and rather stayed at major version 2.
\pkgname{codemod-cli} have $^\wedge2.48.0$ constraint for {simple-git} from $V0.8.7$ to $V0.9.1$.
``updated''=\textcolor{blue}{false} indicates that the package has an outdated version of the dependency in this interval.

At the beginning of row 85, the resolved version of \pkgname{simple-git} $V2.48.0$ was found vulnerable to four vulnerabilities (\texttt{CVE-2022-\allowbreak{}24066}~\cite{CVE-2022-24066}, \texttt{CVE-2022-24433}~\cite{CVE-2022-24433}, \texttt{CVE-2022-25912}~\cite{CVE-2022-25912}, and \texttt{CVE-\allowbreak{}2022-25860}~\cite{CVE-2022-25860}).
In this example, we only consider \texttt{CVE-2022-\allowbreak{}24433} as a vulnerability, which was fixed in $V3.3.0$, released on $\operatorname{2022-03-11}$.
For this reason, we mark the intervals from rows 85-87 as ``remediated'' = \textcolor{red}{false} since \pkgname{codemod-cli} has a vulnerable version of \pkgname{simple-git} even though a fixed version is available.
``remediated''=\textcolor{red}{false} means that the package has an outdated and vulnerable version of the dependency in this interval.
In $V0.9.2$, \pkgname{codemod-cli} bumped the constraint of {simple-git} to $^\wedge3.4.0$, which resulted in $V3.4.0$, a fixed version of the above vulnerability.
So \pkgname{codemod-cli} has the highest available version of \pkgname{simple-git} in this interval (row 88) and so marked ``remediated''=true.

When computing the Time-To-Update (\ttu) for the dependency relation between \pkgname{codemod-cli} and {simple-git}, we sum up the intervals with ``updated''=\textcolor{blue}{false} with exponential weighting.
With the weighting factor, \ttu for \pkgname{codemod-cli} becomes 2.4 days, which is less than naively computing the delta between $\operatorname{2020-07-16}$ to $\operatorname{2022-03-22}$ (rows 1 - 87), which is $614$ days.
Since this period of outdated dependency (rows 1 - 87) occurred in the year 2022, the weighting factor ensures giving less emphasis on that.
Similarly, \ttr for \pkgname{codemod-cli} (rows with ``remediated''=\textcolor{red}{false}) becomes 3.67 days.
Because of the weighting factor, our computed \ttr=3.67 is lower than naively summing up the delta between $\operatorname{2022-03-11}$ to $\operatorname{2022-03-22}$ (rows 85 - 87), which is $11$ days.
Since this period of intervals with ``remediated''=\textcolor{red}{false} is older, these intervals are weighted accordingly in \ttr.
We formally define \mttu and \mttr in Section~\ref{sec:definitions} and explain our design choices.

\subsection{Metrics Definitions}
\label{sec:definitions}
We present a formal definition of our proposed metrics in this section.

\begin{metric}[$\operatorname{\mathbf{Mean-Time-To-Update}}_{dep}$ : $\mathbf{MTTU}_{dep}$]
    \mttu of a package is the weighted aggregated time the package uses an outdated direct dependency version in its lifetime.
\end{metric}
We have described how to calculate the \ttu\ for a \pkgdep relationship in Section~\ref{sec:running-example}.
Formally speaking, \ttu\ of a package $p_i$ with considering only dependency $p_j$  ($<p_i, p_j>$ relationship) is defined in Equation~\ref{eq:mttu-pkgdep}.
\setlength{\abovedisplayskip}{0pt} 
\begin{align}
    {TTU}_{dep}(p_i, p_j) = \frac{\sum_t w_t d_t}{\sum_t w_t}
    \label{eq:mttu-pkgdep}
\end{align}

\mttu for package $p_i$ with $n$ direct dependencies is defined in Equation~\ref{eq:mttu}.
\begin{align}
    {MTTU}_{dep}(p_i) = \frac{\sum\limits_{j=1}^{n} {TTU}_{dep}(p_i, p_j) }{n}
    \label{eq:mttu}
\end{align}
Here, $d_t$ indicates one interval duration with `updated'=\textcolor{blue}{false} for $<p_i, p_j>$ relationship which ends at timestamp $t$.
Also, $w_t = \exp (-\lambda a_t)$ is the weight assigned for interval $d_t$ and $a_t$ is the age of the interval $d_t$.
The age of an interval is the number of days from the end of the interval to our data collection date (2026-05-07).
$\lambda$ is the decaying factor in this weighting function, and we set $\lambda=\frac{\ln(2)}{\tau}$.
In this equation, $\tau$ is the half-life, and we set $\tau=2\ years$ since 2 years is used in literature to assess recent ongoing maintenance~\cite{miller_understanding_2025,li_comparison_2023}.
With this decaying weight, recent intervals approach full weight, while the weight of the older intervals decays exponentially to near zero.
Our use of weighted average is inspired by similar other research~\cite{guzman_emotional_awareness_2013,ajiono_time_series_forecasting_2023,lee2016comparison,sikand_green_ai_2023}.

We considered linear ($w_t = max(a) - a_t + \epsilon$), exponential ($w_t = \exp (-\lambda a_t)$), and inverse ($w_t = \frac{1}{a_t+\epsilon}$) as the choices for the weighting function based on the criteria described by Ulan et al.~\cite{ulan_weighted_2021} for weighted quality scoring for software metrics.
We opted for the exponential weighting function since exponential weighting is more responsive to recent data than linear weighting, and is more configurable and robust than inverse weighting (using $\tau$).
For example, if a development team wants to consider 3 years as an appropriate half-life for their specific case, they can configure the weighted version by changing $\tau=3\ years$.

A vulnerable dependency is a security risk: it gives an attacker a way to exploit the package.
Remediation means the package's developers have released a version that replaces that vulnerable dependency version with a fixed one, closing off that avenue of exploitation.
\mttr is designed to track exactly this property, separately from ordinary (non-security) updates.

\begin{metric}[$\operatorname{\mathbf{Mean-Time-To-Remediate}}_{dep}$ : $\mathbf{MTTR}_{dep}$]
    \mttr of a package is the weighted aggregated time a package uses an outdated and vulnerable direct dependency version in its lifetime.
\end{metric}
\ttr of a package $p_i$ with considering only dependency $p_j$ is defined in Equation~\ref{eq:mttr-pkgdep}.
\begin{align}
    {TTR}_{dep}(p_i, p_j) = \frac{\sum_t w_t d_t}{\sum_t w_t}
    \label{eq:mttr-pkgdep}
\end{align}

\mttr for $n$ direct dependencies is defined in Equation~\ref{eq:mttr}.
\begin{align}
    {MTTR}_{dep}(p_i) = \frac{\sum\limits_{j=1}^{n} {TTR}_{dep}(p_i, p_j) }{n}
    \label{eq:mttr}
\end{align}
Here, $d_t$ indicates one interval duration with `remediated'=\textcolor{red}{false} for $<p_i, p_j>$ relationship which ends at timestamp $t$.
Also, $w_t = \exp (-\lambda a_t)$ is the weight assigned for interval $d_t$ and $a_t$ is the age of the interval $d_t$.

\begin{tcolorbox}[colback=green!5!white,colframe=green!75!black]
    \textbf{Answer to RQ1:}
    Our design of \mttu and \mttr overcomes the limitations of existing dependency update metrics.
\end{tcolorbox}

\section{Empirical Study Methodology}
In this section, we first present vulnerability, package metadata (versions and dependency relations), and the packages' characteristics collection process to apply the metric implementation in the three ecosystems (\rqref{rq:mttu-mttr}).
We then present the statistical tests used to compare \mttu and \mttr across ecosystems and vulnerability severities (\rqref{rq:mttu-mttr}).
After that, we present a statistical testing process to verify if MTTU can be a proxy for MTTR, and how sensitive that verdict is to the metrics' weighting function (\rqref{rq:substitute}).
Lastly, we present correlation tests and regression analysis to understand how package characteristics impact MTTU and MTTR (\rqref{rq:characteristics}).

\subsection{Data Collection}

\noindent\textbf{\textit{Vulnerability Information.}}
We collect the CVE data for our chosen three ecosystems from osv.dev~\cite{osv-dev} for npm, PyPI, and Cargo packages on 2026-05-07.
We rely on OSV since OSV aggregates CVE data from multiple sources (e.g., GitHub Security Advisories, PyPA, GoVulDB)~\cite{osv-data-sources} in one place.
After downloading JSON-formatted CVE data from OSV, we convert it into an SQL table that includes \textit{ecosystem}, \textit{package name}, \textit{CVE ID}, \textit{version where the vulnerability was introduced}, and \textit{version where the vulnerability was fixed}.

\noindent\textbf{\textit{Package Metadata.}}
We collect the package-version history data and dependency information for npm, PyPI, and Cargo packages from deps.dev on 2026-05-07, similar to other previous studies~\cite{lin_untrustide_2024,zahan_software_2023}.
We chose the three ecosystems to have a diverse set of ecosystems, where npm is the largest, PyPI is the oldest, and Cargo is the newest among the major software ecosystems.
In our dataset, we have initially $2{,}603{,}314$ npm, $274{,}720$ PyPI, and $122{,}069$ Cargo packages.
We collected data from deps.dev since it provides all package versions and dependency information for our three chosen ecosystems.
After data collection, we split each \pkgdep relation into multiple intervals.
Since the dependency resolution can be complex and differs across ecosystems, we use deps.dev to perform the dependency resolution allowed by the dependency constraints, with our added requirement (only using the available versions of the dependency before the interval start time).

\noindent\textbf{\textit{Package Characteristics.}}
We examined several characteristics (from Saini et al.~\cite{saini_investigating_2020}) for packages: the number of contributors; the number of dependencies and dependents; the number of version releases; the SourceRank score, and the number of forks and stars.
On 2026-05-07, we downloaded these characteristics of each package from libraries.io to examine their association with a package's \mttuw and \mttrw values.
Libraries.io is used by other research as a data source~\cite{gu_self-admitted_2023,cao_towards_2023,he_migrationadvisor_2021}.
When counting the number of dependencies, we only use the number of dependencies of packages in their latest version.
To ensure construct validity in downloading these characteristics, we manually inspected a sample of packages to verify the characteristics, and we found the downloaded characteristics to be accurate.
Additionally, we computed the number of major versions and package ages for our packages and added them to our analysis.

\subsection{Package Inclusion and Exclusion Criteria}
\label{sec:inclusion-exclusion}
We begin with an initial dataset of $3{,}000{,}103$ ($2{,}603{,}314$ npm, $274{,}720$ PyPI, and $122{,}069$ Cargo) packages collected from deps.dev.
Our first step is to apply two inclusion criteria: \textbf{\textit{(1)}} the package must be at least two years old (operationalized by the difference between the first and last version release); and \textbf{\textit{(2)}} the package must have at least one residual activity (e.g., one version release) in the last two years.
Miller et al.~\cite{miller_understanding_2025} used two years of residual activity followed by two years of no maintenance as criteria to find out the abandoned packages.
Our criteria are inspired by Miller et al., since we want to include the packages that are maintained.
Moreover, ``two years'' is a commonly used standard adopted by other research to measure whether a package is maintained or not~\cite{li_comparison_2023}.
However, our package selection criteria might miss packages that are less than two years old or have had no activity in the last two years (e.g., feature complete packages~\cite{coelho2017modern}), even if they are not abandoned.
We then used our exclusion criteria: a package without any dependencies should be excluded.
Since our metrics characterize packages' dependency update practice, packages without any dependencies do not fit into our study.
The number of packages after these exclusion criteria is \textbf{$163{,}207$ ($117{,}129$ npm, $42{,}777$ PyPI, and $3{,}301$ Cargo packages)}, which will be the final set we use in our analyses.
Out of these $163{,}207$ packages, $22{,}513$ ($17{,}263$ npm, $5{,}158$ PyPI, and $92$ Cargo) packages have at least one vulnerable dependency in their lifetime, supporting our initial motivation of lack of vulnerability data.

\subsection{Metrics Implementation}
After resolving the dependencies and applying our inclusion-exclusion criteria, we store the data in a PostgreSQL database.
Given the large size of the dataset, working directly with the raw data would be impractical. We also create indexes on the most frequently accessed keys to speed up data retrieval for our metrics calculations.

In the database, we compute if the resolved dependency version matches the highest available version of the dependency during each interval and mark each interval accordingly, as shown in the ``updated'' column in Table~\ref{table:running-example}.
Similarly,  we then compute if the resolved dependency version for each interval is vulnerable to any security advisory, even though a fixed version is available, and mark the interval accordingly in the ``remediated'' column in Table~\ref{table:running-example}.

\subsection{Ecosystem and Severity-Stratified Comparison}
\label{sec:methods-ecosystem-stats}
Violin plots show \emph{how} the \mttu and \mttr distributions of npm, PyPI, and Cargo differ, but not whether a difference is statistically and practically meaningful.
We therefore split \rqref{rq:mttu-mttr} into two sub-questions, and we answer each one with statistical tests.

\begin{description}
  \item[RQ2a (ecosystem comparison).] Do npm, PyPI, and Cargo differ in how quickly out-of-date dependencies are updated (\mttu) and how quickly known vulnerable dependencies are remediated (\mttr)? If they differ, how large is the difference in practice?
  \item[RQ2b (severity-stratified comparison).] Within an ecosystem, does remediation speed depend on vulnerability severity, and does the severity/ecosystem relationship hold consistently across ecosystems?
\end{description}

\noindent\textbf{\textit{Data and Error Exclusions.}}
Both sub-questions reuse the \mttu/\mttr values already computed for the full $163{,}207$-package population (Section~\ref{sec:inclusion-exclusion}), under our primary exponential weighting function ($\tau=2$ years).
Recomputing package-level \mttu for this statistical comparison fails for a fraction of packages, e.g., a deps.dev record that could not be re-resolved or had missing data.
We drop these error rows rather than imputing them.
This leaves $n=123{,}318$ packages for the RQ2a \mttu comparison: $98{,}962$ npm, $22{,}042$ PyPI, and $2{,}314$ Cargo.
These counts are $84.5\%$, $51.5\%$, and $70.1\%$ of each ecosystem's full \mttu population, respectively.
PyPI loses the largest share, so we treat this as a limit on how far the RQ2a \mttu comparison generalizes to the full PyPI population.
\mttr recomputation did not lose any packages.
The RQ2a \mttr comparison therefore uses the full $22{,}513$-package population reported in Section~\ref{sec:inclusion-exclusion}: $17{,}263$ npm, $5{,}158$ PyPI, and $92$ Cargo.
For RQ2b, we compute \mttr separately for each severity level of a package.
The unit of analysis in RQ2b is therefore a package--severity pair, and one package can contribute up to four pairs.
RQ2b uses inverse weighting as its primary source, because the per-severity \mttr is available only under inverse weighting and no weighting.
We repeat RQ2b under no weighting as a robustness check.

\noindent\textbf{\textit{Statistical Tests.}}
\mttu and \mttr are heavily right-skewed, and their maxima are two to three orders of magnitude above the median (Section~\ref{sec:results-rq2}).
ANOVA, $t$-tests, and Tukey's HSD assume normal groups with equal variances, so they do not fit these data.
We therefore test for an overall difference across ecosystems (RQ2a) or severities (RQ2b) with the rank-based Kruskal--Wallis $H$-test~\cite{kruskalwallis1952}.
Where this test is significant, we compare each pair of groups with a two-sided Mann--Whitney $U$ test~\cite{mannwhitney1947}, with a Holm--Bonferroni correction~\cite{holm1979} within each set of comparisons.
A set is the 3 ecosystem pairs for RQ2a, and the up to 6 severity pairs within one ecosystem, or the 3 ecosystem pairs within one severity, for RQ2b.

With up to $98{,}962$ packages per group, even small differences become significant.
We therefore also report Cliff's delta~\cite{cliff1993}, a rank-based effect size, computed from the Mann--Whitney statistic as $\delta=2U/(n_1n_2)-1$.
We use the thresholds of Romano et al.~\cite{romano2006}: negligible ($|\delta|<0.147$), small ($<0.33$), medium ($<0.474$), or large ($\geq 0.474$).

The Kruskal--Wallis test can only show that the severity groups differ in some way.
For RQ2b, we therefore also test whether remediation speed follows severity monotonically.
We compute the Spearman rank correlation~\cite{spearman1904} between \mttr and the severity rank (critical$=4>$high$=3>$medium$=2>$low$=1$, from OSV~\cite{osv-dev}), for each ecosystem and pooled.
We implement all tests with \texttt{scipy.stats} and \texttt{statsmodels}.

\subsection{Proxy and Sensitivity Analysis}
\label{subsec:proxy-analysis}
To answer \rqref{rq:substitute}, we split it into two sub-questions.

\begin{description}
  \item[RQ3a (proxy analysis).] Can \mttu be used as a substitute/proxy for \mttr, based on regression, sensitivity, and partial-correlation analyses informed by prior work?
  \item[RQ3b (weighting sensitivity).] Does the answer to RQ3a hold regardless of the choice of weighting function?
\end{description}

\noindent\textbf{\textit{RQ3a: Proxy Analysis.}}
We apply the following analyses to \mttu and \mttr to assess whether one variable (\mttu) can serve as a substitute or proxy for the other (\mttr), following prior work.
Before we describe them, we state what we expect from a proxy.
We do not want to predict the exact \mttr value of a package from its \mttu value.
We want to estimate how a package is likely to remediate vulnerable dependencies from two kinds of information that exist for every package.
The first is its observed dependency update behavior (\mttu).
The second is its context, which we describe with nine characteristics of the package, its repository, and its developers.
These are the same characteristics that we study in \rqref{rq:characteristics}: contributors count, dependents count, forks, SourceRank, stars, dependencies count, version count, major version count, and package age.
\mttu is a useful proxy only if it adds information about \mttr beyond this context.

\noindent\textbf{(1) Covariate-Adjusted Regression with Wald Test.} Several works~\cite{wooldridge2010econometric, mahnken2014evaluating} proposed the use of regression to identify proxies. 
A proxy should have (1) a statistically significant slope ($p_\mathrm{value} <$ 0.05) and (2) residuals with a zero mean, approximately normal distribution, and small variance. 
Montgomery et al.~\cite{montgomery2000measuring} also used (3) a Wald test on the proxy variable coefficient. 

A simple regression of \mttr on \mttu cannot tell us whether \mttu carries its own information about remediation.
Both metrics can depend on the same package context.
For example, a large and active package may update all of its dependencies quickly, whether they are vulnerable or not.
In that case, \mttu and \mttr agree only because both reflect the context, and the context alone would give the same estimate.
Wooldridge~\cite{wooldridge2010econometric} therefore evaluates a proxy together with the observed covariates in the model, because a proxy is useful only for the part of the unobserved target variable by quantifying the degree to which it explains variation not already accounted for by covariates
in the model.
We follow this approach and fit one Ordinary Least Squares (OLS) model per weighting function:
\begin{equation}
\label{eq:proxy-adjusted}
MTTR_i = \beta_0 + \beta_1\, MTTU_i + \sum_{k=1}^{9} \gamma_k\, x_{ik} + \varepsilon_i ,
\end{equation}
where $x_{i1},\dots,x_{i9}$ are the nine characteristics of package $i$.
We keep \mttu and \mttr in days, so that $\beta_1$ has a direct meaning in days.
The nine characteristics are heavily right-skewed counts, so we apply a $\log(1+x)$ transformation to each one to reduce the influence of a few very large packages.
We then standardize each one (z-score), so that all covariates are on the same scale.
The residual variance of \mttr can differ across packages with long-tailed \mttu values, so we use robust HC3 estimation of standard errors that is consistent under heteroscedasticity~\cite{long2000using}.
Diagnostic residual plots suggest that variability in \mttr may increase with \mttu.  
We check multicollinearity with the variance inflation factor (VIF)~\cite{obrien2007caution}.

In practice, $\beta_1$ is the expected difference in \mttr between two packages that have the same nine characteristics but differ by one day in \mttu.
The model thus compares a package with its peers of similar context, not with the whole ecosystem.
We test $H_0: \beta_1 = 0$ with a Wald test on the same adjusted model~\cite{montgomery2000measuring}.
As this test has a single restriction, its statistic is the square of the robust $z$-statistic of $\beta_1$, so it confirms the coefficient test and does not add separate evidence.

A significant coefficient does not show that \mttu explains much of \mttr, especially with tens of thousands of packages.
We therefore measure the explanatory value of \mttu with a hierarchical comparison~\cite{JCohenPCohenSGWestLS2013}.
We first fit the model with the nine covariates only ($R^2_{cov}$).
We then add \mttu and fit the full model ($R^2_{full}$).
The gain $\Delta R^2 = R^2_{full} - R^2_{cov}$ is the share of the total \mttr variance that only \mttu explains.
The partial $R^2$ of \mttu, $(R^2_{full} - R^2_{cov}) / (1 - R^2_{cov})$, is the share of the variance that the context leaves unexplained and that \mttu explains~\cite{JCohenPCohenSGWestLS2013}.
The partial $R^2$ answers our question directly: how much does \mttu tell us about \mttr beyond the context?
We grade $R^2$ as moderate at $0.09$ or above and strong at $0.25$ or above.
These thresholds are the squares of Cohen's medium ($r=0.3$) and large ($r=0.5$) correlations~\cite{cohen1988}.

The regression needs all nine characteristics for each package.
It therefore uses the complete-case sample: the packages that match the characteristic data, have all values present, and have $\mttu \geq 0$ and $\mttr > 0$.
This sample has $22{,}175$ packages under exponential weighting, $19{,}765$ under no weighting, and $19{,}406$ under inverse weighting.
We use the \texttt{statsmodels} library in Python for the regression and the Wald test.

\noindent\textbf{(2) Sensitivity Analysis.} Seltzer~\cite{seltzer2021perilous} proposed sensitivity analysis for proxy analysis, with $GFI$ (Goodness-of-Fit Index), $AGFI$ (Adjusted Goodness-of-Fit Index), and $NFI$ (Normal Fit Index). 
The indices should be $> 0.90$ for an acceptable fit~\cite{marsh1996assessing}. 
We ran the sensitivity analysis using the \texttt{semopy} library in Python on all packages with \mttr.

\noindent\textbf{(3) Partial Correlation.} We also examine the partial Pearson and Spearman correlations between \mttu and \mttr after controlling for the same nine package characteristics, on the same complete-case sample as the regression. 
The regression measures how much \mttu explains in a linear model of \mttr in days.
The partial correlation gives the strength of the remaining association on a standard $-1$ to $1$ scale, and a non-zero partial correlation shows that the two metrics are not independent once the covariates are fixed~\cite{baba2004partial, li2017variable}.
The rank-based (Spearman) version checks whether a monotonic association remains when we do not assume a linear relation or normal residuals~\cite{liu2018covariate}.
We report coefficients, 95\% confidence intervals, and $p$-values.

We do not use ordinary correlation as a proxy criterion because correlation alone does not control for the shared context. 
An ordinary correlation can be high only because both metrics follow package size, popularity, or activity.
The adjusted regression and the partial correlations remove this shared part and show what \mttu adds beyond it.

\noindent\textbf{\textit{RQ3b: Weighting Sensitivity.}}
For analyzing the sensitivity of the weighting mechanisms (which may have an impact on our conclusion from RQ3a), we recompute the \mttu{} and \mttr{} values for our 163k packages with varying weighting functions (exponential [primary], inverse, and no weighting).
Then we reran the RQ3a proxy analysis on each of these data instances.
If our metric is sensitive to any weighting mechanism, the proxy analysis should report contrasting results among different data instances, and vice versa.

\subsection{Package Characteristics Analysis}
\label{sec:characteristics-analysis}
For analyzing nine package-level characteristics in \rqref{rq:characteristics}, we use correlation tests to estimate if there is any correlation between these characteristics and packages' corresponding \mttuw and \mttrw values.
We use the Spearman rank correlation~\cite{spearman1904}, because the characteristics and the two metrics are heavily right-skewed, and a rank-based measure does not assume a linear relation or normal data.
We read the Spearman coefficient as negligible for $|\rho|<0.1$, weak for $0.1\leq|\rho|<0.3$, moderate for $0.3\leq|\rho|<0.5$, and strong for $|\rho|\geq0.5$~\cite{cohen1988}.
Specifically, we test a hypothesis about the association between the following independent variables and their continuous dependent variables (\mttuw and \mttrw):

\textit{($H_1$) Contributors Count.}. We hypothesize that packages with fewer contributors are less likely to have updated dependencies since they have less capacity for maintenance and dependency management.
\textit{($H_2$) Dependents Count}. We hypothesize that packages with fewer dependents are less likely to have updated dependencies.
Fewer dependents may lead to fewer feature updates, fewer bug fixes, and fewer version releases, which in turn may result in less updated dependencies.
\textit{($H_3$) Dependency Count}. We hypothesize that packages with fewer dependencies are more likely to have updated dependencies.
We expect fewer dependencies to be more manageable, and thus, these projects may have more updated dependencies.

\textit{($H_4$) Version Count}. We hypothesize that packages with fewer version releases are less likely to have updated dependencies.
Since these packages have fewer available versions, they may pay less attention to their dependency management.
\textit{($H_5$) Major Version Count}. We hypothesize that packages with fewer major version releases are more likely to have updated dependencies.
Since these packages have fewer major versions to maintain, they may pay more attention to their dependency management.
\textit{($H_6$) Package Age}. We hypothesize that packages with lower ages are more likely to have updated dependencies.
Lower package age may mean the development team is more proactive in dependency management since the package is not old or mature enough.
Lower package age may also mean that the dependencies do not have scope for publishing many newer versions, which in turn may mean more updated dependencies in these packages.

\textit{($H_7$) SourceRank}. The Package SourceRank score indicates the package quality, popularity, and community metrics calculated in \texttt{libraries.io} dataset~\cite{saini_investigating_2020,sun_using_2023}.
This metric depends on several factors, such as the presence of a README file, license, following SEMVER, recent updates, and the number of contributors.
We hypothesize that packages with lower SourceRank scores are less likely to have updated dependencies.
Since these packages are of low quality, they may not have a lot of dependents or contributors, which in turn may result in less updated dependencies.

\textit{($H_8$) Forks Count}. We hypothesize that packages with fewer forks are less likely to have updated dependencies.
Fewer forks may mean fewer people are using and looking into these packages, which in turn may mean less activity, fewer version releases, and fewer updated dependencies in these packages.

\textit{($H_9$) Stars Count}. We hypothesize that packages with fewer stars are less likely to have updated dependencies.
Fewer stars may indicate fewer people are using and looking into these packages, which consequently may suggest less activity, fewer version releases, and fewer updated dependencies.

\subsection{Regression Analysis}
Correlation analysis measures pairwise relationships between two variables, but does not account for potential interactions between multiple variables.
In contrast, a multiple linear regression model estimates the association between each predictor and the outcome while holding the other predictors constant.
Moreover, strong relationships between two variables from correlation analysis might be influenced by the presence of other variables (confounding variables).
The model estimates the conditional association of each package characteristic with the outcome, accounting for the other included characteristics. For \rqref{rq:characteristics}, we fit multiple linear regression models with package characteristics as predictors and either \mttuw or \mttrw as the outcome. Their coefficients and $p_\mathrm{value}$s indicate the direction and statistical significance of the observed associations.

This process will produce a $p_\mathrm{value}$ for each independent variable to indicate whether the relationship between this variable and the outcome is statistically significant.
To control for family-wise type-I error inflation due to testing multiple dependent models (e.g., models that share the same dependent variable) together, we applied the Bonferroni correction~\cite{bonferroni1935} for the $p_\mathrm{value}$ threshold to determine the statistical significance level.
Since we test nine different characteristics, according to Bonferroni correction, the $p_\mathrm{value}$ of a test needs to be $< 0.05/9\ or\ 0.0055$ to be considered as significant ($p < 0.0055$).
Package characteristics with a significant test result are identified as key package characteristics.
We use the \textsf{statsmodels} package of Python to conduct the correlation tests and to build the regression model.

\section{Empirical Study Results}

\subsection{\rqref{rq:mttu-mttr}: How do packages in npm, PyPI, and Cargo differ in MTTU and MTTR?}
\label{sec:results-rq2}

We answer \rqref{rq:mttu-mttr} in two parts.
RQ2a asks whether npm, PyPI, and Cargo differ overall in \mttu\ and \mttr.
RQ2b asks whether remediation speed depends on vulnerability severity.
Both parts use the statistical tests described in Section~\ref{sec:methods-ecosystem-stats}.

\subsubsection{RQ2a: Statistical Ecosystem Comparison}
\label{sec:results-rq2a}

A median or an interquartile range can show \emph{how} npm, PyPI, and Cargo's \mttu/\mttr distributions differ, but it cannot tell us whether that difference is statistically distinguishable, or how large it is in practice.
So we test this directly, on $n=123{,}318$ packages with a valid \mttu\ recomputation ($98{,}962$ npm, $22{,}042$ PyPI, $2{,}314$ Cargo; the remaining packages could not be recomputed, see Section~\ref{sec:methods-ecosystem-stats}) and the full $22{,}513$-package nonzero-\mttr\ population.
Table~\ref{tab:rq2a-descriptive} gives the descriptive statistics for this comparison.

\begin{table*}[t]
  \centering\small
  \caption{\mttu\ (all packages) and nonzero \mttr\ by ecosystem, primary (exponential, $\tau=2$ years) source. Days unless noted.}
  \label{tab:rq2a-descriptive}
  \begin{tabular}{llrrrrr}
    \toprule
    Metric & Ecosystem & $n$ & Median & IQR & Max & \% zero \\
    \midrule
    \mttu & Cargo & 2,314  & 15.7 & [1.3, 39.7]  & 775   & 20.2\% \\
    \mttu & npm   & 98,962 & 18.6 & [4.9, 45.8]  & 2,653 & 10.7\% \\
    \mttu & PyPI  & 22,042 & 26.0 & [11.3, 54.5] & 981   & 5.0\%  \\
    \midrule
    \mttr\ (nonzero) & Cargo & 92     & 26.5 & [10.0, 42.7] & 516 & -- \\
    \mttr\ (nonzero) & npm   & 17,263 & 23.3 & [11.5, 43.3] & 876 & -- \\
    \mttr\ (nonzero) & PyPI  & 5,158  & 25.4 & [14.0, 46.5] & 493 & -- \\
    \bottomrule
  \end{tabular}
\end{table*}

\noindent\textbf{\mttu\ differs significantly across ecosystems, with small-to-negligible effect sizes.}
A Kruskal--Wallis test rejects the null of equal \mttu\ distributions ($H=1381.6$, $p<10^{-300}$).
All three pairwise comparisons remain significant after Holm correction (Table~\ref{tab:rq2a-pairwise}), but Cliff's $\delta$ is negligible for Cargo-vs-npm ($\delta=-0.10$) and small for Cargo-vs-PyPI ($\delta=-0.25$) and npm-vs-PyPI ($\delta=-0.15$).
In other words, the ecosystems are statistically distinguishable but practically close for \mttu\ under this definition.

\noindent\textbf{\mttr\ differs significantly overall, but only npm-vs-PyPI survives pairwise correction, and every effect size is negligible.}
The omnibus test is significant ($H=41.4$, $p=1.0\times10^{-9}$).
After Holm correction, though, only npm-vs-PyPI remains significant ($p=3.8\times10^{-10}$, $\delta=-0.06$, negligible).
Cargo is not distinguishable from either npm or PyPI (Holm $p=0.80$ and $p=0.79$), which is consistent with Cargo's small \mttr\ sample ($n=92$).

\begin{table*}[t]
  \centering\small
  \caption{Pairwise Mann--Whitney $U$ (Holm-corrected) and Cliff's $\delta$ across ecosystems, primary source.}
  \label{tab:rq2a-pairwise}
  \begin{tabular}{llrlrl}
    \toprule
    Metric & Pair & Holm $p$ & Sig.\ ($\alpha=.05$) & Cliff's $\delta$ & Magnitude \\
    \midrule
    \mttu & Cargo--npm  & $<0.001$ & yes & $-0.10$ & negligible \\
    \mttu & Cargo--PyPI & $<0.001$ & yes & $-0.25$ & small \\
    \mttu & npm--PyPI   & $<0.001$ & yes & $-0.15$ & small \\
    \midrule
    \mttr & Cargo--npm  & 0.80 & no  & $0.02$  & negligible \\
    \mttr & Cargo--PyPI & 0.79 & no  & $-0.05$ & negligible \\
    \mttr & npm--PyPI   & $<0.001$ & yes & $-0.06$ & negligible \\
    \bottomrule
  \end{tabular}
\end{table*}

\begin{figure*}[t]
\centering
\includegraphics[width=\linewidth]{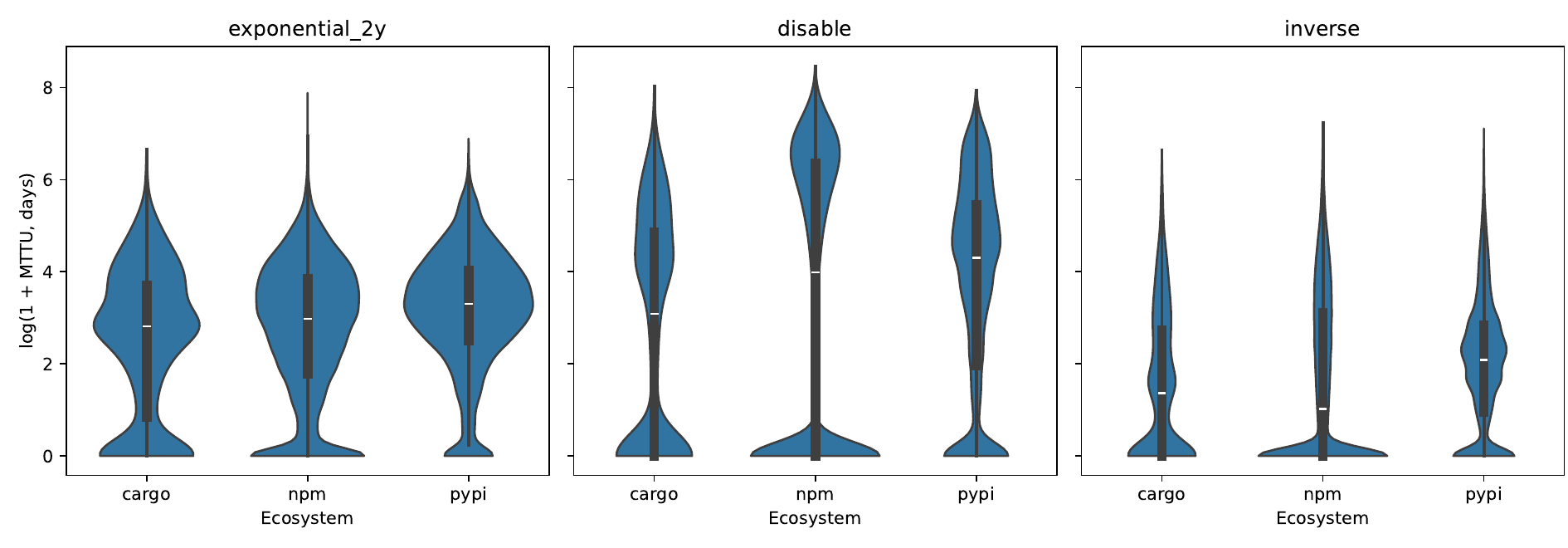}
\caption{\mttu\ by ecosystem, $\log(1+\mttu)$ axis, one panel per weighting scheme (left to right: exponential [primary], no weighting, inverse). Density is truncated at the observed data range.}
\label{fig:rq2a-mttu-violin}
\end{figure*}

\begin{figure*}[t]
\centering
\includegraphics[width=\linewidth]{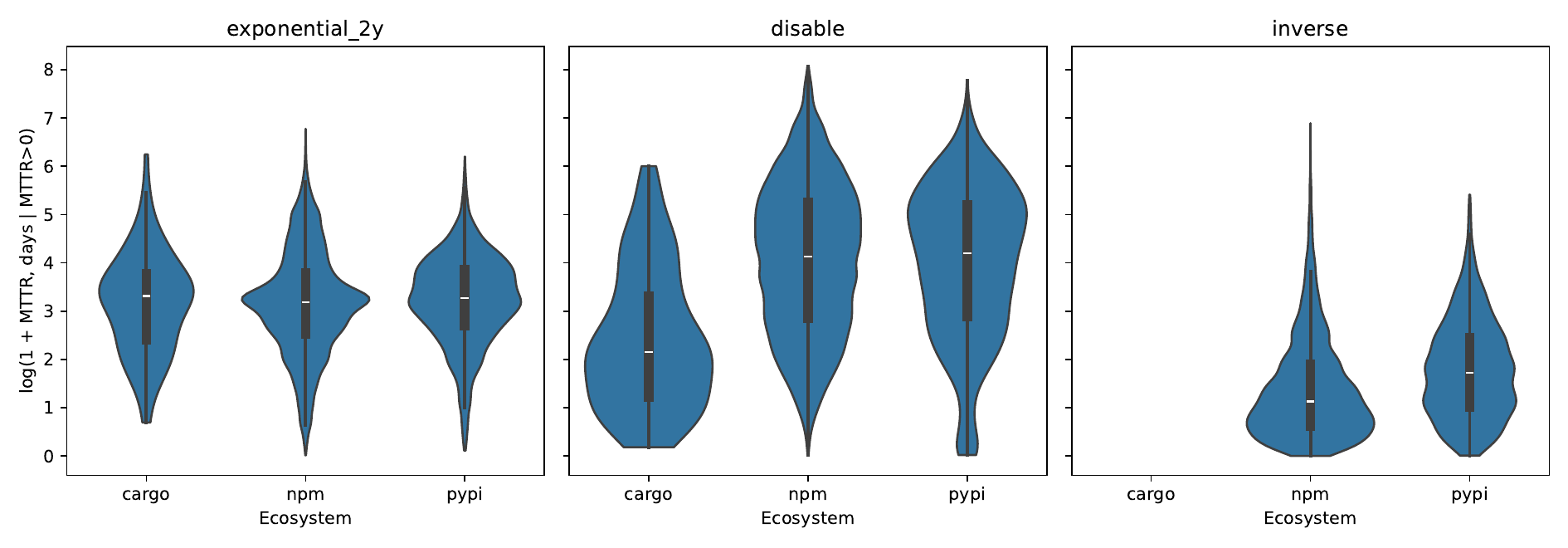}
\caption{Nonzero \mttr\ by ecosystem, $\log(1+\mttr)$ axis, one panel per weighting source (exponential [primary], no weighting, inverse).}
\label{fig:rq2a-mttr-violin}
\end{figure*}

\noindent\textbf{\textit{Robustness Across Weighting Schemes.}}
We recompute \mttu\ under three weighting schemes to check whether the ecosystem difference above is robust to that choice: exponential, no weighting, and inverse.
All three use the same weighted-mean \mttu\ formula defined in Section~\ref{sec:definitions}; they differ only in the weight assigned to each interval, so they are directly comparable to one another and to the primary metric used throughout this paper.
The omnibus difference is significant under all three ($p<10^{-90}$ in every case), so ``ecosystems differ'' is robust.
The \emph{ranking} of which ecosystem is fastest is not robust, however -- it changes with the weighting scheme:
\begin{itemize}
  \item \textbf{Exponential:} Cargo\,$<$\,npm\,$<$\,PyPI ($15.7 < 18.6 < 26.0$ days).
  \item \textbf{No weighting:} the same order, but at much larger magnitudes ($20.8 < 52.9 < 72.8$ days).
  \item \textbf{Inverse:} the order flips to npm\,$<$\,Cargo\,$<$\,PyPI ($1.8 < 2.9 < 7.0$ days).
\end{itemize}
Claims about \emph{which} ecosystem is fastest should therefore be scoped explicitly to the exponential-weighted definition used elsewhere in this paper, not as a general fact about the ecosystems.
We plot this comparison as a violin plot rather than a box plot, since a violin plot shows the same summary statistics (median, spread) while also showing the shape of the distribution, similar to a density plot~\cite{hintzeViolinPlotsBox1998}.
Fig.~\ref{fig:rq2a-mttu-violin} and Fig.~\ref{fig:rq2a-mttr-violin} show this directly.
Each panel uses the same $\log(1+x)$-transformed axis, which compresses the long right tail for readability.
This transform does not change any rank-based test result, since $\log(1+x)$ is strictly increasing over the $x\geq0$ domain \mttu/\mttr live on.

\begin{tcolorbox}[colback=green!5!white,colframe=green!75!black]
    \textbf{Answer to RQ2a:}
    npm, PyPI, and Cargo differ significantly in \mttu, but the effect sizes are small to negligible, and \emph{which} ecosystem is fastest flips across weighting definitions.
    Claims that one ecosystem is ``faster'' should report an effect size and state the weighting definition used.
\end{tcolorbox}

\subsubsection{RQ2b: Severity-Stratified Comparison}
\label{sec:results-rq2b}

Severity-level \mttr\ is only available for the inverse-weighted (primary here) and no-weighting sources.

\begin{table*}[t]
  \centering\small
  \caption{Nonzero per-severity \mttr\ by ecosystem, primary (inverse-weighted) source. Days.}
  \label{tab:rq2b-descriptive}
  \begin{tabular}{llrrrr}
    \toprule
    Ecosystem & Severity & $n$ & Median & IQR & Max \\
    \midrule
    npm  & critical & 1,972  & 1.70 & [0.42, 6.46] & 985.1 \\
    npm  & high     & 9,024  & 1.89 & [0.68, 5.30] & 459.4 \\
    npm  & medium   & 13,117 & 1.92 & [0.75, 5.00] & 816.0 \\
    npm  & low      & 9,771  & 1.29 & [0.57, 2.83] & 251.3 \\
    \midrule
    PyPI & critical & 535   & 3.21 & [1.33, 8.22] & 190.6 \\
    PyPI & high     & 2,139 & 2.87 & [1.26, 6.40] & 205.3 \\
    PyPI & medium   & 2,264 & 3.99 & [1.46, 9.59] & 333.4 \\
    PyPI & low      & 977   & 2.84 & [1.14, 6.78] & 136.8 \\
    \bottomrule
  \end{tabular}
\end{table*}

\noindent\textbf{Only ``low'' severity is clearly faster than the rest; critical, high, and medium are statistically separable but practically indistinguishable.}
Within npm, severities differ overall (Kruskal--Wallis $H=605.6$, $p<10^{-130}$).
After Holm correction, ``low'' has a lower \mttr\ than every other severity.
The effect is small against high ($|\delta|=0.16$) and medium ($|\delta|=0.18$), and negligible against critical ($|\delta|=0.10$, $p<10^{-10}$).
Critical-vs-high and critical-vs-medium are also significant, but both effect sizes are negligible ($|\delta|=0.04$ and $0.05$), so the significance comes from the large samples.
High-vs-medium is not significant (Holm $p=0.27$).
Within PyPI, only high-vs-medium and low-vs-medium are significant after correction, and both effect sizes are negligible ($|\delta|=0.12$ and $0.13$).

\begin{figure*}[t]
\centering
\includegraphics[width=\linewidth]{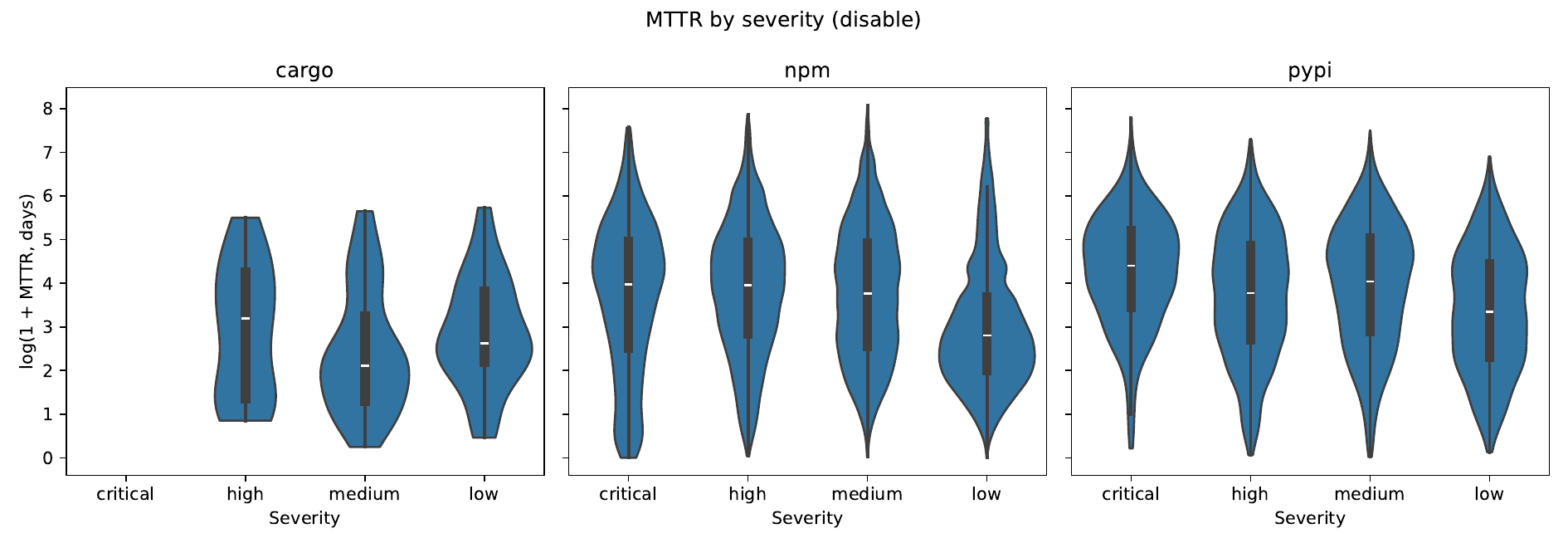}
\caption{Nonzero \mttr\ by severity, $\log(1+\mttr)$ axis, one panel per ecosystem, no-weighting source.}
\label{fig:rq2b-mttr-severity-violin-disable}
\end{figure*}

\noindent\textbf{Severity does not lead to faster remediation speed.}
The Spearman correlation between severity rank (critical$=4>\cdots>$low$=1$) and \mttr\ is small but \emph{positive} for npm ($\rho=0.099$, $p=2.9\times10^{-74}$, $n=33{,}884$) and for the pooled sample ($\rho=0.103$, $p=1.8\times10^{-94}$, $n=39{,}799$).
In other words, more severe vulnerabilities are, if anything, remediated \emph{slightly slower}, not faster.
For PyPI the correlation is near zero and not significant ($\rho=-0.022$, $p=0.095$).
The direction for npm and pooled also holds under the no-weighting robustness source (npm $\rho=0.217$, pooled $\rho=0.205$, both $p<10^{-14}$; Fig.~\ref{fig:rq2b-mttr-severity-violin-disable}).
So this pattern is not an artifact of one weighting choice.

\noindent\textbf{npm and PyPI differ significantly within every severity level, with small-to-medium effect sizes} (Table~\ref{tab:rq2b-pairwise}).
PyPI is consistently slower than npm at every severity (compare medians in Table~\ref{tab:rq2b-descriptive}), most notably at ``low'' ($\delta=-0.33$, medium).

\begin{table*}[t]
  \centering\small
  \caption{npm vs.\ PyPI, within severity: Kruskal--Wallis and Cliff's $\delta$, primary source.}
  \label{tab:rq2b-pairwise}
  \begin{tabular}{lrlrl}
    \toprule
    Severity & Kruskal--Wallis $p$ & Sig. & Cliff's $\delta$ (npm vs.\ PyPI) & Magnitude \\
    \midrule
    critical & $1.8\times10^{-14}$ & yes & $-0.22$ & small \\
    high     & $6.6\times10^{-29}$ & yes & $-0.15$ & small \\
    medium   & $2.6\times10^{-84}$ & yes & $-0.26$ & small \\
    low      & $1.3\times10^{-65}$ & yes & $-0.33$ & medium \\
    \bottomrule
  \end{tabular}
\end{table*}

\noindent \textbf{\textit{Comparison to Prior Work.}}
In contrast to prior research~\cite{kula_developers_2018}, we found that the majority of packages in an ecosystem have a lower \mttr.
The reason behind our metrics encompassing lower \mttr values is that our metrics weight recency, and older dependency update practice has less impact on our metrics.
We also observed a long-tail distribution for both \mttu and \mttr in all three ecosystems.
A significant number of packages lag behind in keeping dependencies updated, which is similar to the observations of Cox et al.~\cite{cox_measuring_2015}.
In short, regularly updating dependencies might not be a widespread practice.
Our observation is similar to vulnerable dependency updates as well.
Even with weighting, some packages took months or longer to remediate known vulnerable dependencies, similar to observations from Kula et al.~\cite{kula_developers_2018}.
This suggests that challenges in keeping dependencies updated are not specific to any ecosystem but rather general in nature.

\begin{tcolorbox}[colback=green!5!white,colframe=green!75!black]
    \textbf{Answer to RQ2b:}
    Remediation speed is not statistically associated with vulnerability severity.
    Only \emph{low}-severity vulnerabilities are remediated faster.
    The severity-speed correlation is weakly \emph{positive}.
    npm packages are remediated faster than PyPI at every severity level.
\end{tcolorbox}

\begin{table}[h]
    \centering
    \footnotesize
    \caption{Proxy Analysis Result (primary, exponential weighting)}
    \label{tab:proxy-result}
    \setlength{\aboverulesep}{0pt}\setlength{\belowrulesep}{0pt}
    \renewcommand{\arraystretch}{1.25}
    \resizebox{\linewidth}{!}{
    \begin{tabular}{llc}
        \toprule
        \rowcolor{tblhead}\textbf{Criterion} & \textbf{Result} & \textbf{Pass?}\\
        \midrule
        \tgroup{3}{(1) Covariate-adjusted regression ($n=22{,}175$)}
        Coefficient $\beta_1$ & $0.751$ (95\% CI: $0.710$--$0.791$), $p<.001$ & \tpass{Yes} \\
        Wald test & $\chi^2(1)=1314.4$, $p<.001$ & \tpass{Yes} \\
        $R^2$ (full model) & $0.349$, strong & \tpass{Yes} \\
        Partial $R^2$ of \mttu & $0.287$, strong (covariates alone: $0.088$) & \tpass{Yes} \\
        \tgroup{3}{(2) Sensitivity analysis (all packages with \mttr)}
        GFI & $1.00$ ($>0.90$) & \tpass{Yes}\\
        AGFI & $1.00$ ($>0.90$) & \tpass{Yes}\\
        NFI & $1.00$ ($>0.90$) & \tpass{Yes}\\
        \tgroup{3}{(3) Partial correlation, nine controls ($n=22{,}175$)}
        Pearson & $r=0.571^{***}$ & \tpass{Yes}\\
        Spearman & $\rho=0.567^{***}$ & \tpass{Yes}\\
        \bottomrule
    \end{tabular}}
    \par\vspace{2pt}\scriptsize{*** $p<.001$. The regression (Equation~\ref{eq:proxy-adjusted}) and the partial correlations control simultaneously for all nine package characteristics on the complete-case sample. The sensitivity analysis uses all packages with \mttr.}
\end{table}

\subsection{\rqref{rq:substitute}: Can MTTU serve as a proxy for MTTR?}
\label{sec:results-rq3}

We answer \rqref{rq:substitute} in two parts.
RQ3a asks whether \mttu can serve as a proxy for \mttr using the proxy criteria described in Section~\ref{subsec:proxy-analysis}.
RQ3b asks whether that answer changes under different weighting functions.
Both parts use the analyses described in Section~\ref{subsec:proxy-analysis}.

\subsubsection{RQ3a: Proxy Analysis}
\label{sec:results-rq3a}

As specified in Section~\ref{subsec:proxy-analysis}, the proxy analysis covered regression, sensitivity, and partial-correlation analyses.
Table~\ref{tab:proxy-result} summarizes the results under the primary exponential weighting.

\noindent\textbf{(1) Covariate-Adjusted Regression with Wald Test.}
The estimated partial regression coefficient was positive ($\hat{\beta_1} = 0.751$, robust $SE = 0.021$, 95\% CI: $0.710$--$0.791$, $n = 22{,}175$).
The Wald test rejected $\beta_1 = 0$ ($\chi^2(1)=(0.751/0.021)^2=1314.4$, $p < .001$).
To interpret the estimated partial regression coefficient, consider two populations of packages that are equivalent in all covariates and differ only by $10$ days in the \mttu measurement. 
The estimated difference in the average \mttr of these two populations is $10\hat{\beta_1} = 7.5$ days.

The coefficient shows the direction of the relation, but not how much a linear function of \mttu explains.
The nine covariates alone explain only 8.8\% of the variance in \mttr ($R^2_{cov} = 0.088$).
When we add \mttu, $R^2$ rises to $0.349$ (adjusted $R^2 = 0.349$), a gain of $\Delta R^2 = 0.261$.
The partial $R^2$ of \mttu is $0.287$.
Thus, \mttu explains 28.7\% of the \mttr variance that the package context leaves unexplained, which is a strong effect under our thresholds.
The context adds almost nothing once \mttu is in the model: on the same sample, \mttu alone gives $R^2 = 0.348$.
Hence, the observed update behavior of a package tells us much more about its remediation behavior than its size, popularity, activity, or age do.

Multicollinearity does not affect this result.
The VIF of \mttu is $1.30$.
Forks and stars have VIF values above 10 (the largest is $13.8$ for forks), so their own coefficients are unstable.
However, when we remove forks from the model, $\beta_1$ changes only from $0.7510$ to $0.7508$.

The $R^2$ of this model is much higher than the $R^2 = 0.032$ of a simple regression on all packages with \mttr.
This change comes mostly from the sample, not from the covariates.
On the complete-case sample, the simple regression of \mttr on \mttu already gives $R^2 = 0.348$.
We use the complete-case sample because the adjustment needs all nine characteristics.
We report the change so that the reader does not attribute the higher $R^2$ to the covariates.

\noindent\textbf{(2) Sensitivity Analysis.} The sensitivity analysis returns $GFI = 1.00, AGFI=1.00$, and $NFI=1.00$. As all three indices meet the acceptable threshold (0.90), this result indicates that \mttuw does provide a good fit to \mttrw.

\noindent\textbf{(3) Partial Correlation.} On packages with \mttr, the partial Pearson correlation controlling simultaneously for contributors count, dependents count, forks, SourceRank, stars, dependencies count, version count, major version count, and package age was $r=0.571$ (95\% CI: $0.56$--$0.58$, $p<.001$, $n=22{,}175$).
The covariate-adjusted Spearman correlation was $\rho=0.567$ (95\% CI: $0.56$--$0.58$, $p<.001$).
Both show a moderate-to-strong positive association that remains after adjustment.
The Spearman result also shows that the association does not depend on a linear relation in days.
Table~\ref{tab:partial-covariates} presents the separate analyses controlling for each covariate one at a time.
No single covariate removes the association: the partial Pearson $r$ stays between $0.582$ and $0.590$.

\noindent\textbf{\textit{Interpretation.}}
All three analyses agree under the primary weighting.
Among packages with a similar context, a package that is slow to update its dependencies is also likely to be slow to remediate its vulnerable dependencies.
This relation is not an artifact of package size, popularity, activity, or age, because the nine characteristics are held fixed.
However, part of the relation is expected by construction, because the two metrics share their intervals (Section~\ref{sec:threats}).
Therefore, \mttuw is a context-adjusted proxy for \mttrw, and the strength of the proxy is an upper estimate.
However, the full model still leaves 65\% of the variance in \mttrw unexplained.
Thus, \mttuw can show which packages are likely to remediate slowly compared to their peers, but it cannot give the exact remediation time of a package.

\begin{table}[t]
    \centering
    \footnotesize
    \caption{Partial correlations between MTTU and MTTR controlling for each covariate individually, using exponential two-year weighting ($n=22{,}175$ for every row).}
    \label{tab:partial-covariates}
    \setlength{\aboverulesep}{0pt}\setlength{\belowrulesep}{0pt}
    \renewcommand{\arraystretch}{1.2}
    \setlength{\tabcolsep}{4pt}
    \begin{tabular}{lcc}
        \toprule
        \rowcolor{tblhead}\textbf{Covariate controlled} & \textbf{Partial Pearson $r$} & \textbf{Partial Spearman $\rho$} \\
        \rowcolor{tblhead} & \textbf{(95\% CI)} & \textbf{(95\% CI)} \\
        \midrule
        Contributors count & $0.588^{***}$ {\scriptsize (0.58--0.60)} & $0.669^{***}$ {\scriptsize (0.66--0.68)} \\
        \rowcolor{tblstripe}Dependents count & $0.590^{***}$ {\scriptsize (0.58--0.60)} & $0.661^{***}$ {\scriptsize (0.65--0.67)} \\
        Forks & $0.589^{***}$ {\scriptsize (0.58--0.60)} & $0.669^{***}$ {\scriptsize (0.66--0.68)} \\
        \rowcolor{tblstripe}SourceRank & $0.588^{***}$ {\scriptsize (0.58--0.60)} & $0.668^{***}$ {\scriptsize (0.66--0.67)} \\
        Stars & $0.589^{***}$ {\scriptsize (0.58--0.60)} & $0.670^{***}$ {\scriptsize (0.66--0.68)} \\
        \rowcolor{tblstripe}Dependencies count & $0.587^{***}$ {\scriptsize (0.58--0.60)} & $0.667^{***}$ {\scriptsize (0.66--0.67)} \\
        Version count & $0.582^{***}$ {\scriptsize (0.57--0.59)} & $0.586^{***}$ {\scriptsize (0.58--0.59)} \\
        \rowcolor{tblstripe}Major version count & $0.588^{***}$ {\scriptsize (0.58--0.60)} & $0.663^{***}$ {\scriptsize (0.66--0.67)} \\
        Package age (days) & $0.586^{***}$ {\scriptsize (0.58--0.59)} & $0.668^{***}$ {\scriptsize (0.66--0.68)} \\
        \midrule
        \rowcolor{tblprimary}\textbf{All nine together} & $\mathbf{0.571}^{***}$ {\scriptsize (0.56--0.58)} & $\mathbf{0.567}^{***}$ {\scriptsize (0.56--0.58)} \\
        \bottomrule
    \end{tabular}
    \par\vspace{2pt}\scriptsize{*** $p<.001$. The same complete-case sample with positive MTTR is used for every row. Each row above the line controls only for the named covariate. The last row controls for all nine at once (Table~\ref{tab:proxy-result}).}
\end{table}

\begin{tcolorbox}[colback=green!5!white,colframe=green!75!black]
    \textbf{Answer to RQ3a:}
    After we control for nine characteristics of the package, its repository, and its developers, \mttuw remains a significant and strong signal of \mttrw (partial $R^2 = 0.287$, partial $r = 0.571$). Thus, \mttuw can serve as a context-adjusted proxy for \mttrw when vulnerability data is unavailable. \MTTU shows which packages are likely to remediate slowly compared to similar packages, but it does not replace \mttrw, as 65\% of the variance in \mttrw remains unexplained.
\end{tcolorbox}

\subsubsection{RQ3b: Weighting Sensitivity}
\label{sec:results-rq3b}

Section~\ref{sec:definitions} explains our reasoning that exponential weighting is more responsive to recent activity than linear weighting, and more configurable than inverse weighting.
That reasoning does not, by itself, tell us whether the alternatives would have changed our proxy-analysis conclusion.
In this sub-question, we explore whether the choice of weighting function and half-life would have changed the conclusion.
We rerun the analyses from Section~\ref{subsec:proxy-analysis} two more times, once with no weighting at all (every point in a package's history counts equally) and once with inverse weighting (recency still matters, but decays more gradually than under our exponential choice).
No weighting is a useful baseline because it tells us whether time-decay weighting matters at all.
Inverse weighting tells us whether our specific decay shape matters, or whether any decay shape would do.
Together with the primary exponential result already reported in RQ3a, this gives us three points across the weighting design space.
This section covers only the weighting-\emph{function} part of RQ3b.
We did not rerun the analysis at different half-lives of the exponential function itself, so half-life sensitivity is still open for future work.

Table~\ref{tab:rq3b-weighting-sensitivity} shows the outcomes of the proxy analyses under each weighting function. The partial-correlation estimates are reported as effect sizes and are not treated as pass/fail tests.

\begin{table}[t]
    \centering\footnotesize
    \caption{Proxy analysis outcome across weighting functions.}
    \label{tab:rq3b-weighting-sensitivity}
    \setlength{\aboverulesep}{0pt}\setlength{\belowrulesep}{0pt}
    \setlength{\tabcolsep}{5pt}
    \renewcommand{\arraystretch}{1.25}
    \begin{tabular}{l>{\columncolor{tblprimary}}ccc}
        \toprule
        \rowcolor{tblhead}\textbf{Criterion} & \textbf{Exponential} & \textbf{None} & \textbf{Inverse} \\
        \rowcolor{tblhead} & \textbf{(primary)} & & \\
        \midrule
        \tgroup{4}{(1) Covariate-adjusted regression}
        Coefficient $\beta_1$ & \tpass{$0.751$} & \tpass{$0.253$} & \tpass{$0.181$} \\
        Wald $\chi^2(1)$ & \tpass{$1314.4$} & \tpass{$1699.0$} & \tpass{$137.5$} \\
        $R^2$ (full model) & \tpass{$0.349$} & \tpass{$0.325$} & \tpart{$0.159$} \\
        Partial $R^2$ of \mttu & \tpass{$0.287$} & \tpass{$0.250$} & \tpart{$0.121$} \\
        Covariates-only $R^2$ & $0.088$ & $0.100$ & $0.044$ \\
        \tgroup{4}{(2) Sensitivity analysis}
        GFI & \tpass{$1.00$} & \tfail{$0.55$} & \tpass{$1.00$} \\
        AGFI & \tpass{$1.00$} & \tfail{$0.10$} & \tpass{$1.00$} \\
        NFI & \tpass{$1.00$} & \tfail{$0.55$} & \tpass{$1.00$} \\
        \tgroup{4}{(3) Partial correlation (nine controls)}
        Pearson $r$ & $0.571^{***}$ & $0.504^{***}$ & $0.364^{***}$ \\
        Spearman $\rho$ & $0.567^{***}$ & $0.519^{***}$ & $0.345^{***}$ \\
        \midrule
        Complete-case $n$ & 22,175 & 19,765 & 19,406 \\
        \bottomrule
    \end{tabular}
    \par\vspace{2pt}\scriptsize{\textcolor{passfg}{\ding{51}}~pass (strong for $R^2$), \textcolor{partfg}{$\sim$}~moderate, \textcolor{failfg}{\ding{55}}~fail ($<0.90$). *** $p<.001$. All coefficients and Wald tests have $p<.001$. The complete-case $n$ applies to the regression and the partial correlations. The sensitivity analysis uses all packages with \mttr.}
\end{table}

\noindent\textbf{\textit{The adjusted association holds no matter which weighting function we use.}}
Under all three weighting functions, the adjusted coefficient of \mttuw is positive and significant, and so is its Wald test.
The size of $\beta_1$ ranges from $0.181$ to $0.751$.
We do not compare these sizes directly, because each weighting function puts \mttu and \mttr on a different scale in days.
The partial $R^2$ does not depend on this scale, so we use it to compare the settings.
The partial $R^2$ of \mttu is strong under exponential weighting ($0.287$) and no weighting ($0.250$), and moderate under inverse weighting ($0.121$).
In every setting, the nine covariates alone explain at most 10\% of the variance in \mttr, and \mttu adds more than they do.
The partial correlations are also positive for all three weighting functions (Pearson $r=0.364$--$0.571$; Spearman $\rho=0.345$--$0.567$).
Thus, the direction of the adjusted association is stable, while its strength depends on the weighting choice.

\noindent\textbf{\textit{Sensitivity-analysis fit depends on which weighting function we use.}}
The sensitivity-analysis fit indices are not stable.
Under exponential and inverse weighting, $GFI$, $AGFI$, and $NFI$ are all $1.00$, well above the $0.90$ threshold for a good fit.
Under no weighting, the same three indices drop to $0.55$, $0.10$, and $0.55$, well below it.
So the same underlying data can look like a good fit or a bad fit, purely depending on whether we apply time-decay weighting at all.

\noindent\textbf{\textit{Overall verdict.}}
Only the primary exponential weighting passes every criterion.
No weighting has a strong adjusted regression but fails the sensitivity indices.
Inverse weighting passes the sensitivity indices but has only a moderate adjusted regression.
The weighting choice therefore changes how strong the proxy is, but not whether \mttuw carries information about \mttrw beyond the package context.
A claim about the strength of \mttuw as a proxy for \mttrw should specify the weighting function used.

\begin{tcolorbox}[colback=green!5!white,colframe=green!75!black]
    \textbf{Answer to RQ3b:}
    Under every weighting function, \mttuw remains a significant, positive signal of \mttrw after we control for the nine package characteristics (partial $R^2$ from $0.121$ to $0.287$). Only our primary exponential weighting passes all proxy criteria. The other two weighting functions give a weaker or less consistent proxy.
\end{tcolorbox}

\subsection{\rqref{rq:characteristics}: How do package characteristics influence MTTU and MTTR?}

\begin{figure}
    \centering
    \includegraphics[width=\columnwidth]{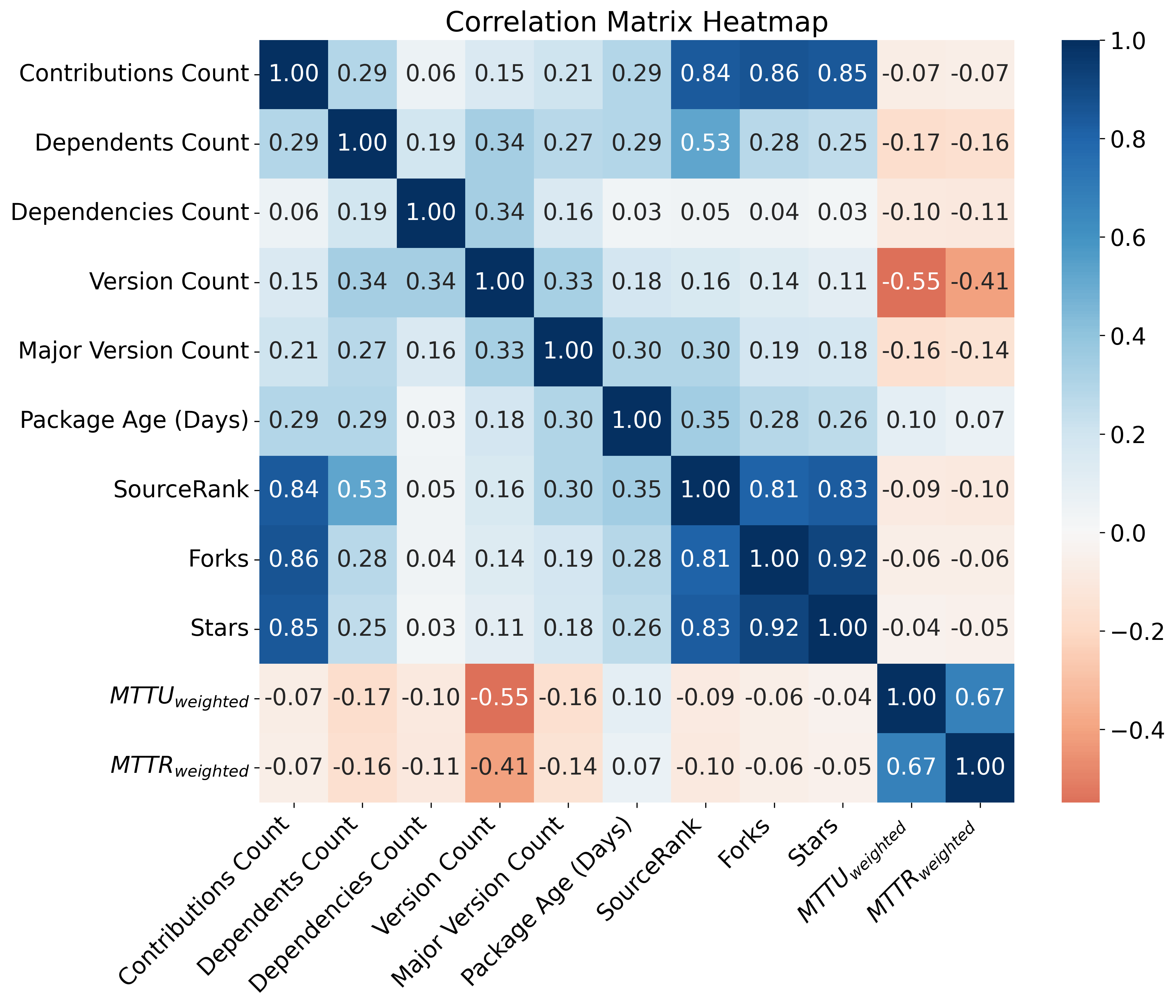}
    \caption{Spearman correlation matrix on packages' characteristics and \mttu and \mttr values.}
    \label{fig:corrw}
\end{figure}

\noindent\textbf{\textit{4a Correlation Analysis.}}
We present the Spearman correlation matrix with the nine packages' characteristics and \mttuw and \mttrw data in Figure~\ref{fig:corrw}.
We label each correlation with the thresholds in Section~\ref{sec:characteristics-analysis}.
Except for version count, the correlations with \mttuw and \mttrw are weak or negligible.

\textit{Contributors Count ($H_1$)}:
We hypothesized that packages with fewer contributors would be less likely to maintain updated dependencies.
The contributors count has negligible correlations with \mttuw ($\rho=-0.07$) and \mttrw ($\rho=-0.07$), suggesting a limited influence on dependency management.

\textit{Dependents Count ($H_2$)}:
We hypothesized that packages with fewer dependents are less likely to have updated dependencies.
The dependents count has weak negative correlations with \mttuw ($\rho=-0.17$) and \mttrw ($\rho=-0.16$).
This suggests that the number of dependents plays a limited role in determining the timeliness of dependency updates.

\textit{Dependency Count ($H_3$)}:
We hypothesized that packages with fewer dependencies would have more updated dependencies.
The dependency count has weak negative correlations with \mttuw ($\rho=-0.10$) and \mttrw ($\rho=-0.11$), which is the opposite of the predicted direction.
This implies that the number of dependencies in a package has little association with dependency management efficiency.

\textit{Version Count ($H_4$)}:
We hypothesized that packages with fewer version releases are less likely to have updated dependencies.
The version count has a strong negative correlation with \mttuw ($\rho=-0.55$) and a moderate negative correlation with \mttrw ($\rho=-0.41$).
This supports our hypothesis that packages with more releases tend to have lower measured dependency update and remediation times in our sample.

\textit{Major Version Count ($H_5$)}:
We hypothesized that packages with fewer major version releases are more likely to have updated dependencies.
However, the major version count has weak negative correlations with \mttuw ($\rho=-0.16$) and \mttrw ($\rho=-0.14$), which is the opposite of the predicted direction.
This suggests that the number of available major versions to maintain has a limited impact on how quickly dependencies are updated.

\textit{Package Age ($H_6$)}:
We hypothesized that packages with lower ages are more likely to have updated dependencies.
Package age has a weak positive correlation with \mttuw ($\rho=0.10$) and a negligible correlation with \mttrw ($\rho=0.07$).
This suggests that package age has little impact on having updated dependencies.
This also strengthens our intuition of creating weighted versions, \mttuw and \mttrw, to eliminate the potential age-sensitivity of \mttu and \mttr.

\textit{SourceRank ($H_7)$}.
We hypothesized that low quality packages (e.g., packages with lower SourceRank scores) are less likely to have updated dependencies.
The results give limited support for this hypothesis: SourceRank has a negligible correlation with \mttuw ($\rho=-0.09$) and a weak correlation with \mttrw ($\rho=-0.10$).

\textit{Forks Count ($H_8$)}:
We hypothesized that packages with fewer forks are less likely to have updated dependencies.
Fork count and contributors count have a strong positive correlation ($\rho=0.86$), indicating that packages with more forks tend to have more contributors. This association also suggests potential multicollinearity between these predictors in the regression models.
However, the forks count has negligible correlations with \mttuw ($\rho=-0.06$) and \mttrw ($\rho=-0.06$), suggesting that having more forks does not directly translate into faster dependency updates.

\textit{Stars Count ($H_9$)}:
We hypothesized that packages with fewer stars are less likely to have updated dependencies.
Star count and contributors count have a strong positive correlation ($\rho=0.85$), indicating that packages with more stars tend to have more contributors. This association also suggests potential multicollinearity between these predictors in the regression models.
However, the star count has negligible correlations with \mttuw ($\rho=-0.04$) and \mttrw ($\rho=-0.05$), implying that more stars do not directly indicate faster dependency updates.

Overall, except for version count, our results indicate weak or negligible correlations between the tested factors and \mttu and \mttr.
These findings suggest that the analyzed factors may not strongly influence how updated dependencies are, and further exploration of other variables or non-linear relationships is recommended.
Table~\ref{tab:h1-h9-summary} summarizes all nine hypotheses at a glance: the direction we predicted, the observed Spearman correlation with \mttuw/\mttrw, and whether the hypothesis held.

\begin{table*}[t]
    \centering\small
    \caption{Summary of hypotheses $H_1$--$H_9$: predicted direction, observed Spearman correlation with \mttuw/\mttrw, and whether the hypothesis held.}
    \label{tab:h1-h9-summary}
    \setlength{\aboverulesep}{0pt}\setlength{\belowrulesep}{0pt}
    \renewcommand{\arraystretch}{1.25}
    \begin{tabular}{clccl}
        \toprule
        \rowcolor{tblhead} & \textbf{Characteristic} & \textbf{Predicted} & \textbf{Correlation (\mttuw/\mttrw)} & \textbf{Held?} \\
        \midrule
        $H_1$ & Contributors count & $-$ & $-0.07$ / $-0.07$ & \tfail{No (negligible)} \\
        \rowcolor{tblstripe}$H_2$ & Dependents count & $-$ & $-0.17$ / $-0.16$ & \tpart{Partially (weak)} \\
        $H_3$ & Dependency count & $+$ & $-0.10$ / $-0.11$ & \tfail{No (opposite)} \\
        \rowcolor{tblstripe}$H_4$ & Version count & $-$ & $-0.55$ / $-0.41$ & \tpass{Yes (moderate)} \\
        $H_5$ & Major version count & $+$ & $-0.16$ / $-0.14$ & \tfail{No (opposite)} \\
        \rowcolor{tblstripe}$H_6$ & Package age & $+$ & $\phantom{-}0.10$ / $\phantom{-}0.07$ & \tpart{Partially (weak)} \\
        $H_7$ & SourceRank & $-$ & $-0.09$ / $-0.10$ & \tpart{Partially (weak)} \\
        \rowcolor{tblstripe}$H_8$ & Forks count & $-$ & $-0.06$ / $-0.06$ & \tfail{No (negligible)} \\
        $H_9$ & Stars count & $-$ & $-0.04$ / $-0.05$ & \tfail{No (negligible)} \\
        \bottomrule
    \end{tabular}
    \par\vspace{2pt}\scriptsize{\textcolor{passfg}{\ding{51}}~held, \textcolor{partfg}{$\sim$}~partially held, \textcolor{failfg}{\ding{55}}~did not hold. Predicted $-$: we expect a higher value of the characteristic to go with a lower (better) \mttuw/\mttrw; predicted $+$: the reverse.}
\end{table*}

\begin{tcolorbox}[colback=green!5!white,colframe=green!75!black]
    \textbf{Answer to RQ4a:}
    Except for version count, the package characteristics have weak or negligible Spearman correlations ($|\rho|\leq0.17$) with \mttuw and \mttrw.
    Packages with a higher version count have lower \mttuw ($\rho=-0.55$) and \mttrw ($\rho=-0.41$).
\end{tcolorbox}

\noindent\textbf{\textit{4b Regression Analysis (\mttuw As Dependent Variable).}}
Interpreting the multilinear regression model's characteristics, we found 0.039 as the $R^2$ value when \mttuw was the dependent variable.
$R^2$ value 0.039 indicates that 3.9\% variation in the dependent variable can be explained by the model.
Although the model is statistically significant ($p_\mathrm{value} <$  0.0055 from the F-test), a lower $R^2$ value indicates that other factors, beyond the nine independent variables, also substantially impact \mttuw.
The model's F-statistic is 547.7 ($p<0.0001$).
This indicates that at least one of the independent variables (or predictors) has a non-zero relationship with \mttuw.

We then look into the coefficients and $p_\mathrm{value}$s associated with each of the independent variables.
The coefficient indicates the expected change in the dependent variable for a one-unit change in one independent variable while holding other independent variables constant.
The $p_\mathrm{value}$ associated with the t-statistic indicates whether the independent variable is statistically significant in explaining the variation in the dependent variable.
A lower $p_\mathrm{value}$ ($<$ 0.0055) would indicate statistical significance.

Some independent variables have positive coefficients, indicating that increases in those variables are associated with increases in \mttuw, holding the other predictors constant.
Independent variables with positive coefficients are contributors count (0.0025), dependents count ($5.3\times10^{-5}$), forks (0.0002), and package age (0.0114).
However, the rest of the coefficients are small, which indicates these independent variables do not have a significant impact on \mttuw.

We also found that some independent variables have negative coefficients.
Negative coefficients indicate that increases in those independent variables are associated with decreases in \mttuw, holding the other predictors constant.
Independent variables with negative coefficients are SourceRank (-2.2260), stars (-0.0002), dependencies count (-0.0493), and major versions count (-0.4766).
The dependencies count result runs counter to our hypothesis ($H_3$): we expected fewer dependencies to mean a lower (better) \mttuw, i.e., a positive coefficient, but both the correlation analysis above and this regression find the opposite direction.
A possible explanation is that packages with more dependencies tend to be more actively maintained overall (e.g., a higher version count, per $H_4$), and that general activity dominates any extra burden from having more dependencies to track; we leave confirming this mechanism to future work.
SourceRank, which indicates package quality and popularity, shows moderate negative coefficients.
Higher SourceRank indicates an overall better project.
The model predicts that if SourceRank for a package increases by 1 (moving to a ``better'' rank), \mttuw drops by 2 days.
Moreover, packages that have released more major versions tend to have a lower \mttuw.
Finally, independent variables, except dependent count and forks, were found to be statistically significant ($p_\mathrm{value} <$ 0.0055).
\begin{tcolorbox}[colback=green!5!white,colframe=green!75!black]
    \textbf{Answer to RQ4b:}
    Older packages have a slightly higher \mttuw, but the model explains only 3.9\% of the variance.
    Popular and better quality packages (with higher SourceRank scores) are more likely to have lower \mttuw.
\end{tcolorbox}

\noindent\textbf{\textit{4c Regression Analysis (\mttrw As Dependent Variable).}}
For \mttrw as the dependent variable, the multilinear regression model results in 0.026 as the $R^2$ value.
$R^2$ = 0.026 indicates that the model explains 2.6\% of the variation in \mttr.
A lower $R^2$ value indicates that other unmodeled factors play an important role in determining \mttrw, which is similar to our observations in \mttuw.
The model's F-statistic is 65.4 ($p<0.0001$).
This indicates the model is statistically significant and at least one of the independent variables (or predictors) has a non-zero relationship with \mttrw.

We then look into the coefficients and $p_\mathrm{value}$s associated with each of the independent variables.
We found that independent variables, except for the dependent count, forks, and stars, are statistically significant ($p_\mathrm{value} <$ 0.0055) in modeling \mttrw.

Package age has a small positive coefficient (0.0091) and is statistically significant: older packages are associated with slightly higher \mttrw.
Forks (coefficient 0.0008) and dependents count (coefficient $2.3\times10^{-5}$) also have positive coefficients, but neither association is statistically significant at our corrected threshold. Coefficient magnitudes depend on each predictor's scale, so their small numerical values alone do not indicate small practical associations.
Additionally, the major version count (coefficient -0.3453), dependencies count (coefficient -0.1213), contributors count (coefficient -0.0107), and SourceRank (coefficient -1.3284) show a moderate negative effect on \mttrw.
As with \mttuw\ above, the dependencies count direction runs counter to $H_3$'s prediction; we do not have a confirmed explanation for this and leave it to future work.
This observation is otherwise similar to our previous observation with \mttuw.
According to the model's prediction, an increase in SourceRank by 1 (i.e., a higher SourceRank indicates a better package) would result in a 1-day reduction in \mttrw.

\begin{tcolorbox}[colback=green!5!white,colframe=green!75!black]
    \textbf{Answer to RQ4c:}
    Similar to \mttuw, older packages have a slightly higher \mttrw, but the model explains only 2.6\% of the variance.
    Popular and better quality packages (with higher SourceRank scores) are more likely to have lower \mttrw.
    Packages with higher major versions are more likely to have lower \mttrw.
\end{tcolorbox}

\section{Discussions And Implications}
\label{sec:discussion}

\noindent\subsection{Practical Implications For Developers}
\noindent\textbf{\textit{When \mttu And \mttr Should Be Used.}}
0.49\% npm, 0.83\% PyPI, and 0.05\% Cargo packages have directly depended on at least one vulnerable package in their lifetime, and we can compute $MTTR_{dep}$ for these packages.
Whereas we do not know the vulnerable dependency update practice for the majority of the ecosystem (99.51\% npm, 99.17\% PyPI, and 99.95\% Cargo packages).

For this majority, \rqref{rq:substitute} shows that \mttu is a context-adjusted proxy for \mttr.
After we control for nine characteristics of the package, its repository, and its developers, \mttuw still explains 28.7\% of the remaining variance in \mttrw under our primary weighting.
In contrast, these characteristics alone explain only 8.8\%.
In practical terms, among packages with a similar context, packages that are slow to update dependencies also tend to be slow to update vulnerable dependencies.
This finding suggests that improving general dependency update practices will likely also improve vulnerable dependency update practices.

The proxy has a clear limit.
The adjusted model leaves 65\% of the variance in \mttrw unexplained, so \mttuw does not give a day-for-day estimate of \mttrw.
We therefore recommend the following use:
\begin{enumerate}
    \item Use \mttrw when vulnerability data is available for the package.
    \item When \mttrw is not available, compare the \mttuw of the package with the \mttuw of packages that have a similar context (e.g., similar popularity, size, release activity, and age).
    \item Treat a package whose \mttuw is high compared to these peers as likely to remediate vulnerable dependencies slowly, and review it more closely.
\end{enumerate}
Developers should use \mttuw to rank and screen candidate dependencies, not to predict the exact remediation time of one package.

\noindent\textbf{\textit{Use Floating-Minor With Regular Major Updates.}}
Packages that update their dependencies quickly also tend to remediate quickly, so developers should treat dependency updates as routine maintenance, for example with scheduled updates or automated update notifications.
A constraint that always allows the latest version keeps \mttu and \mttr lowest, but it can bring breaking changes~\cite{ochoa_breaking_2022,raemaekers_semantic_2014,venturini_i_2023}.
A pinned constraint allows no automatic updates, so \mttu starts to grow as soon as the dependency releases a new version, even a patch release.
\emph{A mixed strategy, with automatic minor and patch updates and manual major updates, is therefore a practical way to keep \mttu and \mttr low}.
This recommendation relies on developers following SEMVER correctly~\cite{pinckney_large_2023,li_large-scale_2023,dietrich_dependency_2019}.

\noindent\textbf{\textit{Dependency Selection Criteria.}}
\rqref{rq:characteristics} found that version count and major version count are associated with lower \mttuw/\mttrw: packages with more version releases have lower values (Spearman $\rho=-0.55$/$-0.41$), and packages with more major version releases also have lower values (regression coefficient $-0.4766$/$-0.3453$).
When choosing between otherwise-equivalent dependencies, developers may treat a longer release history as one descriptive signal of maintenance activity, rather than as evidence that release count itself causes faster dependency updates or remediation.
SourceRank also correlates with lower \mttuw/\mttrw, though only weakly ($-0.09$/$-0.10$), so it is a much less reliable signal for this purpose on its own.

\noindent\subsection{Practical Implications For Researchers}
Prior work shows that developers hesitate to update dependencies due to a fear of breaking changes, a lack of awareness or knowledge about available updates, and sometimes a lack of motivation to invest time in updates~\cite{derrKeepMeUpdated2017}.
These human factors are the likely reasons behind the long tail of \mttu and \mttr data.
Our findings also support prior research, revealing that packages with many contributors or higher dependents count do not have a better \mttu or \mttr, in contrast to our hypothesis.
Even packages with thousands of dependents are not guaranteed timely updates for their own dependencies.
Researchers should focus more on exploring the human factors to uncover ways to balance the cost and benefit of dependency updates.
In addition, both multilinear regression models in \rqref{rq:characteristics} present an $R^2$ value of less than 10\%.
A lower $R^2$ value indicates that other unmodeled factors substantially affect \mttu and \mttr.
The proxy analysis in \rqref{rq:substitute} points to the same conclusion from another angle.
The nine characteristics explain only 8.8\% of the variance in \mttrw, but adding \mttuw raises this share to 34.9\%.
Thus, the observed update behavior of a package carries much more information about its remediation behavior than its static characteristics do.
\textit{Researchers who model security responsiveness should include behavioral measures such as \mttu, not only package characteristics.}
Researchers who propose a proxy should also control for the shared context and report the partial $R^2$ of the proxy, because a significant coefficient alone does not show how much the proxy explains~\cite{JCohenPCohenSGWestLS2013}.
The development team dynamics, the use of pinning and floating, the cost and efforts needed in testing, and the size of the codebase could be such possible unmodeled factors.
\textit{Researchers should explore further if such other unmodeled factors influence \mttu and \mttr}.

\noindent\textbf{Version Count and Endogeneity.}
The moderate association between version count and \mttuw/\mttrw does not by itself show that releasing more versions causes faster dependency updates or remediation. 
By design, our metric allows a floating version constraint to resolve to a newer or fixed dependency release without requiring a new release of the package.
Version count is therefore not a necessary route to a lower metric value (faster dependency updates or remediation). 
In practice, the association may reflect developers' greater tendency to publish package releases rather than rely on floating constraints to keep dependencies up-to-date and remediated. 
However, because package releases also define interval boundaries in our metric construction, the current cross-sectional analysis cannot rule out all mechanical coupling or other sources of endogeneity. 
To isolate the mechanism, further research needs a longitudinal analysis with a time order, or a sensitivity analysis that varies the interval construction.

\noindent\textbf{\textit{Metrics' Measurement Should Report A Test And An Effect Size.}}
With tens of thousands of packages, nearly any difference between ecosystems is statistically significant (Section~\ref{sec:results-rq2a}).
Researchers should therefore report an effect size, such as Cliff's $\delta$, to separate a practical difference from an effect of the large sample.
They should also state the weighting definition behind a ranking, because the \mttu ranking of the ecosystems changes across our three weighting functions.
Finally, the weak \emph{positive} correlation between severity and \mttr under inverse weighting (Section~\ref{sec:results-rq2b}) contradicts the common assumption that severe vulnerabilities are remediated faster, and only the direction of the effect, not the Kruskal--Wallis test, shows it.

\noindent\subsection{Practical Implications For Tool Builders}
\noindent Wermke et al.~\cite{wermkeAlwaysContributeBack2023} found that developers select dependencies with numbers that are quick to access, such as downloads, GitHub stars, and the time since the last release.
Our metrics give one such number per package.
\textit{Tool builders can therefore add \mttu and \mttr to dependency management tools and to security risk assessment tools, such as the OpenSSF Scorecard.}

\noindent\subsection{Gaming Metrics}
While it might seem that using a loose floating constraint and bumping dependency constraint periodically could ``game'' our metrics, the underlying outcome is that a package spends minimal time with outdated dependency versions and ensures that security fixes are adopted automatically.
By doing so, this strategy reduces the risk of exploitation from using an outdated dependency version with a known security vulnerability, such as log4Shell~\cite{everson_log4shell_2022,rahman_less_2024}.
In short, trying to game our metrics by adopting floating version constraints is essentially reducing this attack vector for a package.
However, this approach also comes with a tradeoff.
This strategy might make the packages susceptible to malicious package updates (xz-incident~\cite{williams_research_2025}).
Since floating version constraints allow automatic updates, they can also facilitate propagating malicious package updates.

\section{Threats to Validity}
\label{sec:threats}

\noindent \textbf{\textit{External Validity.}}
The main external validity threat is the generalizability of our results to characterize other ecosystems.
While each ecosystem possesses unique features that might not directly correlate with those we studied, we believe the insights gained should also be broadly applicable to other ecosystems.

\noindent \textbf{\textit{Internal Validity.}}
We use the security advisory dataset from OSV.dev, which may not be comprehensive.
If an advisory is published but not included in the OSV dataset, that may impact our results.
Additionally, we do not consider whether the vulnerable dependency version is exploitable or reachable~\cite{cisa-vex} from the package.
By design, \mttu and \mttr treat all vulnerabilities equally, regardless of CVSS score or severity; using severity as a weighting factor in the metrics themselves would be an interesting future work.
Separately from the metric design, our severity-stratified analysis (Section~\ref{sec:results-rq2b}) does examine whether remediation speed varies with severity, using the critical/high/medium/low bands reported by our OSV-derived vulnerability source; that analysis has its own limitations worth noting here: Cargo could not be included at all due to a lack of severity-tagged remediation data, and we use Spearman rank correlation as a practical substitute for a dedicated ordinal trend test (Jonckheere--Terpstra), which is not available in the \texttt{scipy}/\texttt{statsmodels} stack used throughout this paper.
After downloading the data from deps.dev, we manually checked 20 packages' versions and relations with the public package registries and found that the data is accurate.
Moreover, each package manager has its own way of handling dependency resolutions, and for the dependency resolutions, we rely on the Open Source Insights~\cite{depsdev} resolved version data.
Our analysis omits package versions not adhering to SEMVER rules, a conservative choice to enable a more rigorous analysis.
In addition, Open Source Insights dependency resolution fails in some cases (e.g., missing timestamp), and we mark those as warnings in our dataset.
We do not calculate update metrics for those packages, and we argue that this might have a very small impact on our results.
In addition, we only consider runtime dependencies in our \mttu and \mttr\ analysis and omit dev and optional dependencies.
Additionally, some dependencies might be more important than others depending on the context; however, we treat each dependency equally since modeling dependencies' importance is out of the scope of our study.

\noindent \textbf{\textit{Conclusion Validity.}}
Our proxy analysis is cross-sectional, so it shows an association between \mttu and \mttr, not a causal effect.
We control for nine measured characteristics, but unmeasured factors, such as team practices or the use of automated update tools, can affect both metrics.
In addition, part of the association between \mttu and \mttr is expected by construction.
Both metrics use the same intervals.
An interval with ``remediated''=false is also always an interval with ``updated''=false, because a vulnerable resolved version cannot be the highest available version when a fixed version exists.
The covariate adjustment does not remove this shared part, so the partial $R^2$ of \mttu is an upper estimate of the information that \mttu adds about \mttr.
Two analyses can separate the two metrics: compute \mttu only from the intervals that were never vulnerable, or compute \mttu before a cut-off date and \mttr only for the advisories published after it.
We leave these analyses to future work.
The covariate-adjusted regression and the partial correlations use the complete-case sample of about 20 thousand packages per weighting function, while the sensitivity analysis uses all packages with \mttr.
The simple regression of \mttr on \mttu gives a much higher $R^2$ on the complete-case sample than on the full sample, so the strength of the proxy can differ for packages outside this sample.
Forks and stars are collinear (VIF above 10), so we do not interpret their own coefficients in this model.
The VIF of \mttu stays between $1.08$ and $1.30$, and its coefficient is almost unchanged when we remove forks.
Finally, our moderate and strong $R^2$ thresholds follow common conventions~\cite{cohen1988}, and other thresholds can change the labels, but not the reported values.

\section{Conclusion and Future Works}
In this study, we introduced two dependency update metrics, \mttu and \mttr, to quantify the updatedness of dependencies in open-source software packages.
Our large-scale empirical analysis across the npm, PyPI, and Cargo ecosystems found that \mttu is a context-adjusted proxy for \mttr when vulnerability information is unavailable.
After we control for nine characteristics of the package, its repository, and its developers, \mttu still explains 28.7\% of the remaining variance in \mttr under our primary weighting, and its association with \mttr holds under all three weighting functions.
Thus, \mttu shows which packages are likely to remediate slowly compared to similar packages, but it does not replace \mttr when \mttr is available.
Our statistical comparison further showed that the three ecosystems differ significantly in \mttu, but with small or negligible effect sizes, and the fastest ecosystem changes with the weighting definition.
For \mttr, only npm and PyPI differ significantly, and the effect size is negligible.
Under inverse weighting, remediation speed is only weakly associated with vulnerability severity, and more severe vulnerabilities are, if anything, remediated slightly slower.
Furthermore, our statistical analysis highlighted the relationships between package characteristics and dependency update behavior, providing actionable insights for developers, maintainers, and software supply chain researchers.
Future research can explore additional factors influencing dependency update practices, such as incorporating vulnerability severity directly into the metrics' weighting function, organizational policies, or developer incentives.
Future research can also test the context-adjusted proxy on packages outside our complete-case sample and over time, for example by checking whether a high \mttu today is followed by slow remediation of vulnerabilities that are disclosed later.
Expanding this analysis to other ecosystems, combining with transitive dependencies, and incorporating qualitative insights from developers could further refine our understanding of dependency updatedness.

\section*{Data Availability}
The code and data for the analysis in this paper are available in our replication package in Zenodo~\cite{zenodo-artifact}.

\bibliographystyle{IEEEtran}
\bibliography{references,websites,zotero}

\end{document}